\documentclass[twocolumn,amsmath,showpacs,eqsecnum,twoside]{revtex4}

\usepackage{amssymb}
\usepackage{graphicx}

\begin{document}

\title{Square-gradient mechanism of surface scattering \\ in quasi-1D rough waveguides
}

\author{F.~M.~Izrailev}
\affiliation{Instituto de F\'{\i}sica, Universidad Aut\'{o}noma
         de Puebla, \\
         Apartado Postal J-48, Puebla, Pue., 72570, M\'{e}xico}

\author{N.~M.~Makarov}
\thanks{On sabbatical leave from Instituto de Ciencias,
        Universidad Aut\'{o}noma de Puebla,
        Priv. 17 Norte No. 3417, Col. San Miguel
        Hueyotlipan, Puebla, Pue., 72050, M\'{e}xico}

\affiliation{A.~Ya.~Usikov Institute for Radiophysics and
             Electronics, Ukrainian Academy of Science,
             12 Proskura Street, 61085 Kharkov, Ukraine}

\author{M.~Rend\'{o}n}
\affiliation{Facultad de Ciencias de la Electr\'{o}nica,
         Universidad Aut\'{o}noma de Puebla,
         Puebla, Pue., 72570, M\'{e}xico}

\date{\today}

\begin{abstract}
We develop a perturbative approach that allows one to study the
surface scattering in quasi-1D waveguides with rough boundaries.
Our approach is based on the construction of an effective ``bulk''
potential of a very complicated structure. The detailed analysis
of this potential reveals that apart from the well known terms
considered in the previous studies, one should keep specific terms
that depend on the square of the derivative of the boundaries. As
was found, in spite of an apparent smallness of these {\it
square-gradient scattering} (SGS) terms, there is a physically
important region of parameters, in which they can not be
neglected. Our approach also demonstrates that the contribution of
the SGS-mechanism of scattering strongly depends on the type of
the rough boundaries (uncorrelated, symmetric or antisymmetric).
\end{abstract}


\maketitle

\section{Introduction}
\label{sec-Intro}

Wave scattering in surface-corrugated guiding systems remains a
quite interesting problem, both from the theoretical viewpoint and
due to many applications ranging from the sound and
electromagnetic propagation, to the conduction of electrons in
mesoscopic wires. In spite of many important results obtained to
date (see, e.g., Refs.~\cite{Konrady74,BFb79,McGM84,TJM86,
TrAsh88,IsPuzFuks9091,KMYa91,Tatar93,MeyStep94,MakMorYam95,LunKyReiKr96,BratRash96,
MT98,SFMY99,MT01,LunMenIz01}), there is a number of open questions
to be resolved. One of such problems is to understand the
interplay between different mechanisms of the scattering that are
due to principally different terms in an effective scattering
potential constructed with the use of various perturbative
approaches.

In the recent papers \cite{IzMkRn-PRB-06} it was shown that apart
from the well known terms in the potentials appearing in the first
order perturbation theory, one should keep some specific terms
that were neglected due to their seeming smallness. It was
demonstrated that these terms, although formally belonging to the
second order terms, can play a decisive role in the scattering in
some region of parameters. Since these terms are proportional to
the square of the derivative of surface profiles, they can be
associated with the {\it square-gradient scattering} (SGS)
mechanism.

In contrast with Ref.\cite{IzMkRn-PRB-06} where a particular case
of one rough surface was considered, below we develop a general
approach by concentrating our attention on quasi-one dimensional
waveguides with two corrugated surfaces that are either
statistically uncorrelated or correlated. Our main interest is in
the expression for the {\it attenuation length} $L_n$ (also known
as the scattering length, or the total mean free path),
corresponding to specific $n$-th propagating mode. We assume that
the total number $N$ of these modes (or conducting channels) is
large, $N \gg 1$, and that the lower and upper surface profiles
(boundaries) of such quasi-1D waveguides are described by two
random functions $\xi_{\downarrow}(x)$ and $\xi_{\uparrow}(x)$,
respectively.

In what follows, we consider three configurations of a particular
interest:
\begin{enumerate}
\item
The waveguide with the {\it uncorrelated} boundaries,
$\xi_{\downarrow}(x)$ and $\xi_{\uparrow}(x)$.
\item
The waveguide with the {\it antisymmetric} boundaries,
\begin{equation}\label{CCBcase}
\xi_{\downarrow}(x)=\xi_{\uparrow}(x)=\xi(x).
\end{equation}
\item
The waveguide with the {\it symmetric} boundaries,
\begin{equation}\label{SBcase}
-\xi_{\downarrow}(x)=\xi_{\uparrow}(x)=\xi(x).
\end{equation}
\end{enumerate}

In our approach we assume that due to the multiple scattering of a
travelling wave from the rough boundaries, the longitudinal wave
number of an $n$-th propagating mode can be written as $k_n+\delta
k_n$, where $k_n$ is its unperturbed value,
\begin{equation}\label{k+dk}
\delta k_n=\gamma_n+i(2L_n)^{-1}.
\end{equation}
The real part $\gamma_n$ is responsible for a roughness-induced
correction to the phase velocity of a given mode. As is known, the
shift $\gamma_n$ does not change the transport properties of a
disordered system. Therefore, our further analysis shall be focused
on the attenuation length $L_n$ only.

In order to properly take into account the surface irregularities we
use the approach developed in Ref.~\cite{IzMkRn-PRB-06}.
Specifically, by means of the canonical transformation we formally
reduce the problem of the {\it surface} scattering to the {\it bulk}
one described by the Hamiltonian $\widehat H$,
\begin{equation}\label{H0-U}
\widehat H=\widehat H^{(0)}+\widehat U
\end{equation}
with the effective potential $\widehat U$. It should be stressed
that such a transformation does not mean an equivalence between
the surface and bulk scattering. The very point is that the
obtained ``bulk'' Hamiltonian $\widehat H$ is of very specific
form that does not allow to use standard random matrix approaches
based on a complete randomness in effective potentials.

The paper is organized as follows. In Sec.~\ref{sec-problem} we
start with the general expressions describing the waveguide with
any form of rough boundaries. In Sec.~\ref{sec-DysonEq} we write
the Dyson-type equation and derive approximate expressions for the
surface scattering potential in the form, most useful for our
perturbative approach. In Sec.~\ref{sec-Xi-tilde} we obtain the
fluctuating part of the potential for three configurations of a
waveguide. In Sec.~\ref{sec-averageGF} we present a brief
description of the formalism for obtaining the average Green's
function developed in Ref.~\cite{IzMkRn-PRB-06}. Here we also
specify the conditions of applicability of our method. In
Sec.~\ref{sec-Ln-An} we perform a detailed analysis of the
attenuation length for three specific configurations of the
waveguide. In Sec.~\ref{sec-num-analysis} we report the data that
illustrates our analytical results and make clear the role of
various terms in the expression for the scattering potential. In
this Section we also discuss the so-called ``repulsion''
phenomenon discovered in Ref.~\onlinecite{MnLnIz2004}, in view of
its relation to the square-gradient scattering mechanism. A brief
comparison of the three configurations, as well as our concluding
remarks, are presented in Sec.~\ref{Sec-Conclusion}.

\section{The waveguide with general rough
boundaries}
\label{sec-problem}

In this section we derive the general expressions describing
quasi-1D waveguides with any type of the boundaries. Then, we
present them in the form, most suitable for our analysis of three
particular waveguide configurations.

Specifically, we consider the standard model of an open waveguide
of the average width $d$, stretched along the $x$ axis. The lower
and upper surfaces of the waveguide are assumed to be described,
respectively, by the profiles $z=\sigma\xi_{\downarrow}(x)$, and
$z=d+\sigma\xi_{\uparrow}(x)$. Here $\sigma$ is the
root-mean-square roughness height that is assumed to be identical
for both boundaries. In other words, the waveguide occupies the
region
\begin{equation}\label{Gen}
-\infty\leq x\leq\infty, \qquad
\sigma\xi_{\downarrow}(x)\leq z\leq d+\sigma\xi_{\uparrow}(x)
\end{equation}
of the $(x,z)$-plane. The fluctuating width $w(x)$ is defined by
\begin{equation}\label{Gen-d(x)}
w(x)=d+\sigma[\xi_{\uparrow}(x)-\xi_{\downarrow}(x)], \qquad
\langle w(x)\rangle =d\,,
\end{equation}
where the random functions $\xi_{i}(x)$ (with
$i=\downarrow,\uparrow$) describe the roughness of the
corresponding boundary.

In what follows we consider the model for which both of the
boundaries are statistically identical. They are assumed to be
described by a statistically homogeneous and isotropic Gaussian
random processes with the zero mean and unit variance,
\begin{subequations}\label{Gen-xi}
\begin{eqnarray}
&\langle\xi_{i}(x)\rangle=0,\,\,\,\,\,\,
\langle{\xi_{i}}^2(x)\rangle=1, \\
&\langle\xi_{i}(x)\xi_{i}(x')\rangle={\cal W}(x-x').
\end{eqnarray}
\end{subequations}
Here the angular brackets stand for the statistical averaging over
different realizations of the surface profiles $\xi_i(x)$. We also
assume that the binary correlator ${\cal W}(x)$ decreases on the
scale $R_c$, with the normalization, ${\cal W}(0)=1$.

For three configurations we analyze below, the cross-correlations
are as follows. In the first case of the uncorrelated boundaries
we obviously have,
\begin{equation} \label{UB-BinCorr-xi}
\langle\xi_{\downarrow}(x)\xi_{\uparrow}(x')\rangle=0.
\end{equation}
The waveguide with the antisymmetric boundaries implies:
\begin{equation} \label{CCB-BinCorr-xi}
\langle\xi_{\downarrow}(x)\xi_{\uparrow}(x')\rangle={\cal W}(x-x').
\end{equation}
As for the symmetric boundaries case, the opposite boundaries hold
the correlation equality:
\begin{equation} \label{SB-BinCorr-xi}
\langle\xi_{\downarrow}(x)\xi_{\uparrow}(x')\rangle=-{\cal W}(x-x').
\end{equation}

The {\it roughness-height power} (RHP) spectrum $W(k_x)$ is defined
by:
\begin{equation}\label{FT-W}
W(k_x)=\int_{-\infty}^{\infty}dx\exp(-ik_xx)\,{\cal W}(x).
\end{equation}
Since ${\cal W}(x)$ is an even function of $x$, its Fourier
transform (\ref{FT-W}) is an even, real and non-negative function
of $k_x$. The RHP spectrum has the maximum at $k_x=0$ with
$W(0)\sim R_c$, and decreases on the scale $R_c^{-1}$.

In order to analyze the surface scattering problem we shall employ
the method of retarded Green's function ${\cal G}(x,x';z,z')$.
Specifically, we start with the Dirichlet boundary-value problem
\begin{subequations}\label{Gen-GFP1}
\begin{eqnarray}
&&\left(\frac{\partial^2}{\partial x^2}+
\frac{\partial^2}{\partial z^2}+k^2\right){\cal G}(x,x';z,z')
\nonumber\cr\\
&&=\delta(x-x')\delta(z-z'),
\label{Gen-Geq1}\\[6pt]
&&{\cal G}(x,x';z=\sigma \xi_{\downarrow}(x),z')=0 \nonumber \\
&&{\cal G}(x,x';z=d+\sigma \xi_{\uparrow}(x),z')=0.
\label{Gen-BC1}
\end{eqnarray}
\end{subequations}
Here $\delta(x)$ and $\delta(z)$ are the Dirac delta functions. The
wave number $k$ is equal to $\omega/c$ for an electromagnetic wave
of the frequency $\omega$ and TE polarization, propagating through a
waveguide with perfectly conducting walls. For an electron quantum
wire, $k$ is the Fermi wave number within the isotropic Fermi-liquid
model.

\subsection{Effective surface scattering potential,
$\widehat U$}
\label{subsec-ScatteringPotential}

The equation (\ref{Gen-Geq1}) does not contain any scattering
potential, therefore, the perturbation is hidden in the boundary
conditions (\ref{Gen-BC1}). In this case it is convenient to
perform the canonical transformation to new coordinates,
\begin{subequations}\label{Gen-CoorTr}
\begin{eqnarray}
x_{new}&=&x_{old}, \\[6pt]
z_{new}&=&\frac{d\, [z_{old}-\sigma\xi_{\downarrow}(x_{old})]}
{w(x_{old})}\notag \\
&=&\frac{d\, [z_{old}-\sigma\xi_{\downarrow}(x_{old})]}
{d+\sigma[\xi_{\uparrow}(x_{old})-\xi_{\downarrow}(x_{old})]},
\end{eqnarray}
\end{subequations}
in which both boundaries are flat. In accordance with
Eqs.~(\ref{Gen-CoorTr}), the transformations for the first and
second derivatives with respect to the old coordinates are given
by (below for convenience we drop the subscript $new$ for $x$ and
$z$):
\begin{subequations}\label{Gen-d1Tr}
\begin{eqnarray}
\frac{\partial}{\partial x_{old}}&=&\frac{\partial x}{\partial
x_{old}}\frac{\partial}{\partial x}+\frac{\partial z}{\partial
x_{old}}\frac{\partial}{\partial z} \nonumber \\
&=&\frac{\partial}{\partial x}-
\left[\frac{\sigma\xi'_{\uparrow}(x)}{w(x)}z
+\frac{\sigma\xi'_{\downarrow}(x)}{w(x)}(d-z)\right]
\frac{\partial}{\partial z}\,, \nonumber\\[6pt] \\
\frac{\partial}{\partial z_{old}}&=&\frac{\partial x}{\partial
z_{old}}\frac{\partial}{\partial x}+\frac{\partial z}{\partial
z_{old}}\frac{\partial}{\partial z} \nonumber \\
&=&\frac{d}{w(x)}\frac{\partial}{\partial z}\, ,
\end{eqnarray}
\end{subequations}
\begin{subequations}\label{Gen-d2Tr}
\begin{eqnarray}
\label{Gen-d2xTr}
\frac{\partial^2}{\partial x_{old}^2}&=&
\left\{\frac{\partial}{\partial x}-
\left[\frac{\sigma\xi'_{\uparrow}(x)}{w(x)}z
+\frac{\sigma\xi'_{\downarrow}(x)}{w(x)}(d-z)\right]
\frac{\partial}{\partial z} \right\} \nonumber \\
&\times& \left\{
\frac{\partial}{\partial x}-
\left[\frac{\sigma\xi'_{\uparrow}(x)}{w(x)}z
+\frac{\sigma\xi'_{\downarrow}(x)}{w(x)}(d-z)\right]
\frac{\partial}{\partial z}\right\} \nonumber \\[6pt] \\
\label{Gen-d2zTr}
\frac{\partial^2}{\partial z_{old}^2}&=&\frac{d^2}{w^2(x)}
\frac{\partial^2}{\partial z^2}\,.
\end{eqnarray}
\end{subequations}
Correspondingly, we introduce the canonically conjugate Green's
function,
\begin{equation}
{\cal G}_{new}(x,x';z,z')=\frac{\sqrt{w(x)w(x')}}{d}\,
{\cal G}_{old}(x,x';z,z').
\end{equation}
Here the prefactor before ${\cal G}_{old}$ is due to the Jacobian
of the transformation to new variables. As a result, we arrive at
the equivalent boundary-value problem governed by the equation,
\begin{subequations}\label{Gen-GFP4}
\begin{eqnarray}
&&\Bigg(\frac{\partial^2}{\partial x^2}+
\frac{\partial^2}{\partial z^2} +k^2\Bigg)
\,{\cal G}(x,x';z,z')\nonumber\\[6pt]
&&-\widehat{U}(x,z)\,{\cal G}(x,x';z,z')
=\delta(x-x')\delta(z-z'),\qquad\label{Gen-Geq4}\\[6pt]
&&{\cal G}(x,x';z=0,z')=0,\nonumber\\
&&{\cal G}(x,x';z=d,z')=0.
\label{Gen-BC4}
\end{eqnarray}
\end{subequations}
Here we also omitted the subscript $new$ for ${\cal
G}_{new}(x,x';z,z')$. As a result of the transformation, we have
formally reduced the problem of surface scattering to that of the
bulk one specified by the effective  potential $\widehat{U}(x,z)$,
\begin{eqnarray} \label{Gen-U}
&&\widehat{U}(x,z)=\bigg[1-\frac{d^2}{w^2(x)}\bigg]
\frac{\partial^2}{\partial z^2} \nonumber \cr \\
&&+ \frac{\sigma}{w(x)} \Bigg\{\bigg[\xi_{\uparrow}''(x)
+2\xi_{\uparrow}'(x)\frac{\partial}{\partial x}\bigg]
\bigg[\frac{1}{2}+z\frac{\partial}{\partial z}\bigg]
\nonumber \cr \\
&&-\bigg[\xi_{\downarrow}''(x)+
2\xi_{\downarrow}'(x)\frac{\partial}{\partial x}\bigg]
\bigg[\frac{1}{2}+(z-d) \frac{\partial}{\partial z}\bigg]
 \Bigg\}\nonumber \cr \\
&&-\frac{\sigma^2}{w^2(x)}\Bigg\{{\xi_{\uparrow}'}^2(x)
\bigg[\frac{3}{4}+3z\frac{\partial}{\partial z}
+z^2\frac{\partial^2}{{\partial z}^2}\bigg]\nonumber \cr \\
&&+{\xi_{\downarrow}'}^2(x)
\bigg[\frac{3}{4}+3(z-d)\frac{\partial}{\partial z}
+(z-d)^2\frac{\partial^2}{{\partial z}^2}\bigg] \nonumber \cr \\
&&-\xi_{\downarrow}'(x)\,\xi_{\uparrow}'(x)\bigg[\frac{3}{2}+3(2z-d)
\frac{\partial}{\partial z}+2z(z-d)\frac{\partial^2}{{\partial z}^2}
\bigg]  \Bigg\}.\nonumber \cr \\
\end{eqnarray}
Note that the prime over the function $\xi_i(x)$ denotes a
derivative with respect to $x$. It is important to stress that
Eqs.~(\ref{Gen-GFP4}) and (\ref{Gen-U}) are {\it exact} ones, and
valid for any form of the surface profiles $\xi_{\downarrow}(x)$
and $\xi_{\uparrow}(x)$.

As is expected, the general expression (\ref{Gen-U}) reduces to
that obtained previously in Ref.\cite{IzMkRn-PRB-06} for a
particular case of one rough boundary, if one substitutes
$\xi_{\downarrow}(x)=0$ and $\xi_{\uparrow}(x)=\xi(x)$.

\section{Dyson Equation}
\label{sec-DysonEq}

Here we start with the non-local Dyson equation in
$k_x$-representation,
\begin{eqnarray}
&&g(k_x,k_x')=2\pi\delta(k_x-k_x')\,g_0(k_x)
\nonumber\cr\\
&&+\,g_0(k_x)\int_{-\infty}^{\infty}\frac{dq_x}{2\pi}\,
\Xi(k_x,q_x)g(q_x,k_x') \label{g-DE}
\end{eqnarray}
with $g_0(k_x)=k_z\cot(k_zd)$ being the unperturbed pole-factor.
In this representation the effective surface scattering potential
$\Xi(k_x,k'_x)$ is given by the expression,
\begin{eqnarray}
&&\Xi(k_x,k'_x)=\int_{-\infty}^{\infty}dx_1\int_0^ddz_1\,
\exp\left(-ik_xx_1\right)
\frac{\sin(k_zz_1)}{k_z}
\nonumber\cr\\
&&\times \widehat U(x_1,z_1)
\exp\left(ik'_xx_1\right)\frac{\sin(k'_zz_1)}{k'_z}\,.
\label{Gen-Xi-exact}
\end{eqnarray}
Here $k_x$ and $k_z$ are the lengthwise and transverse wave numbers,
\begin{equation}\label{kz}
k_z=k_z(k_x)=\sqrt{k^2-k_x^2},\qquad k_z'=k_z(k_x').
\end{equation}
Their unperturbed eigenvalues are $k_n$,
\begin{equation}\label{kn}
k_n=\sqrt{k^2-(\pi n/d)^2},\qquad n=1,2,3,\ldots,N_d.
\end{equation}
and $\pi n/d$, respectively. The total number $N_d$ of propagating
waveguide modes (conducting channels that have real values of $k_n$)
is determined by the integer part $[\ldots]$ of the ratio $kd/\pi$,
\begin{equation}\label{Nd}
N_d=[kd/\pi].
\end{equation}

After the substitution of the expression (\ref{Gen-U}) for
$\widehat{U}$ in Eq.~\eqref{Gen-Xi-exact}, we realize that
$\Xi(k_x,k'_x)$ consists of three groups of terms of a different
nature. The first one has the factor $[1-d^2/w^2(x_1)]$, the
second and third groups have the factors $\sigma/w(x_1)$ and
$\sigma^2/w^2(x_1)$, respectively. One needs to note that while
itself the kernel $\Xi(k_x,k'_x)$ is Hermitian, the latter two
parts are non-Hermitian ones. In order to present each part of
$\Xi(k_x,k'_x)$ in an Hermitian form, we perform the integration
by parts for the term containing $\xi_{\downarrow}''(x_1)$ and
$\xi_{\uparrow}''(x_1)$ in the second group. As a result, we
arrive at the final {\it exact} form  for the perturbation
potential,
\begin{eqnarray}
&&\Xi(k_x,k'_x)=\int_{-\infty}^{\infty}dx_1\int_0^ddz_1\,
\exp\left(-ik_xx_1\right) \frac{\sin(k_zz_1)}{k_z} \nonumber\cr\\
&&\times \Biggl(
\bigg[1-\frac{d^2}{w^2(x_1)}\bigg] \frac{\partial^2}{\partial z_1^2}
+ \frac{\sigma}{w(x_1)}\, [i(k_x+k_x')] \nonumber \cr \\
&&\times \bigg\{\xi_{\uparrow}'(x_1)\bigg[\frac{1}{2}
+z_1\frac{\partial}{\partial z_1}\bigg]
-\xi_{\downarrow}'(x_1)\bigg[\frac{1}{2}+(z_1-d)
\frac{\partial}{\partial z_1}\bigg]\bigg\}\nonumber \cr \\
&&- \frac{\sigma^2}{w^2(x_1)}\bigg\{{\xi_{\uparrow}'}^2(x_1)
\bigg[\frac{1}{4}+2z_1\frac{\partial}{\partial z_1}
+z_1^2\frac{\partial^2}{{\partial z_1}^2}\bigg]\nonumber \cr \\
&&+ {\xi_{\downarrow}'}^2(x_1)
\bigg[\frac{1}{4}+2(z_1-d)\frac{\partial}{\partial z_1}
+(z_1-d)^2\frac{\partial^2}{{\partial z_1}^2}\bigg] \nonumber\cr\\
&&- \xi_{\downarrow}'(x_1)\,\xi_{\uparrow}'(x_1)\bigg[\frac{1}{2}
+2(2z_1-d)\frac{\partial}{\partial z_1} \nonumber\cr\\
&&+ 2z_1(z_1-d)\frac{\partial^2}{{\partial z_1}^2}
\bigg] \bigg\} \Bigg)
\exp\left(ik'_xx_1\right)\frac{\sin(k'_zz_1)}{k'_z}
\label{Gen-Xi-exactF}
\end{eqnarray}
This equation has a peculiar structure, very useful for a further
analysis. The kernel written in this form consists also of three
groups of terms, however, now they can be associated with
different scattering mechanisms.

Since we are interested in the averaged Green's function, we have
to calculate the binary correlator of $\Xi(k_x,k'_x)$. Therefore,
in order to avoid very cumbersome calculations, it is reasonable
to make a simplification of $\Xi(k_x,k'_x)$ that does not destroy
the Hermitian structure of each group of terms. Taking also into
account that our main interest is in the case of small surface
corrugations $\sigma\ll d$, we can do the following: expand the
factor $1-d^2/w^2(x_1)\approx
2\sigma[\xi_{\uparrow}(x_1)-\xi_{\downarrow}(x_1)]/d$ in the first
term of Eq.~(\ref{Gen-Xi-exactF}) and to put $w(x_1)\approx d$ in
all the others. In such a way, we derive the following {\it
approximate} expression for the surface scattering potential,
\begin{eqnarray}\label{Gen-Xi-Approx}
&&\Xi(k_x,k'_x)\approx\int_{-\infty}^{\infty}dx_1\int_0^ddz_1\,
\exp\left(-ik_xx_1\right) \frac{\sin(k_zz_1)}{k_z} \nonumber\cr\\
&&\times \Biggl(\frac{2\sigma}{d}\, [\xi_{\uparrow}(x_1)
-\xi_{\downarrow}(x_1)]\frac{\partial^2}{\partial z_1^2}
+ \frac{\sigma}{d}\, [i(k_x+k_x')] \nonumber \cr \\
&&\times \bigg\{\xi_{\uparrow}'(x_1)\bigg[\frac{1}{2}
+z_1\frac{\partial}{\partial z_1}\bigg]
-\xi_{\downarrow}'(x_1)\bigg[\frac{1}{2}+(z_1-d)
\frac{\partial}{\partial z_1}\bigg]\bigg\}\nonumber \cr \\
&&- \frac{\sigma^2}{d^2}\bigg\{{\xi_{\uparrow}'}^2(x_1)
\bigg[\frac{1}{4}+2z_1\frac{\partial}{\partial z_1}
+z_1^2\frac{\partial^2}{{\partial z_1}^2}\bigg]\nonumber \cr \\
&&+ {\xi_{\downarrow}'}^2(x_1)
\bigg[\frac{1}{4}+2(z_1-d)\frac{\partial}{\partial z_1}
+(z_1-d)^2\frac{\partial^2}{{\partial z_1}^2}\bigg] \nonumber\cr\\
&&- \xi_{\downarrow}'(x_1)\,\xi_{\uparrow}'(x_1)\bigg[\frac{1}{2}
+2(2z_1-d)\frac{\partial}{\partial z_1} \nonumber\cr\\
&&+ 2z_1(z_1-d)\frac{\partial^2}{{\partial z_1}^2}
\bigg] \bigg\} \Bigg)
\exp\left(ik'_xx_1\right)\frac{\sin(k'_zz_1)}{k'_z}
\end{eqnarray}

The expression (\ref{Gen-Xi-Approx}) contains three types of
scattering terms of a different physical meaning: the terms
depending on the roughness height $\sigma\xi_{i}(x_1)$, derivative
of the roughness height $\sigma\xi_{i}'(x_1)$, and square of the
derivative of the roughness height $\sigma^2{\xi'_{i}}^2(x_1)$. In
this connection one can speak about the {\it amplitude}, {\it
gradient} and {\it square-gradient scattering} (SGS) mechanisms.
Note that these terms are separately associated with each of two
surfaces. Apart from this, there is a specific group of terms that
are due to the inter-correlations between two surfaces. In what
follows we refer to these terms depending on the product
$\sigma^2\xi'_{\downarrow}(x) \xi'_{\uparrow}(x)$, as to the {\it
gradient inter-product scattering} (GIS) mechanism.

It should be stressed that the GIS-terms, due to their
proportionality to $\sigma^2$, were neglected in previous studies
of the transport properties of surface-disordered waveguides. A
similar situation occurs for the SGS-terms that have been analyzed
only recently in Ref.~\cite{IzMkRn-PRB-06,IzMakRen05}. As we show
below, all these terms can give the main contribution in the
scattering, although formally they may be treated as negligible
ones.

We start our analysis with the presentation of the kernel as a sum
of its average and fluctuating parts, i.e.,
\begin{equation}\label{Xi=<Xi>+tildeXi}
\Xi(k_x,k'_x)=\langle\Xi(k_x,k'_x)\rangle+\widetilde\Xi(k_x,k'_x).
\end{equation}
It can be tested that the average, $\langle\Xi(k_x,k'_x)\rangle$,
contributes only to the real part $\gamma_n$ of the complex
renormalization $\delta k_n$ of the lengthwise wave number $k_n$
(see Eq.~(\ref{k+dk})), and therefore, does not change static
transport properties of the surface disordered waveguide. Thus, we
will omit it because our interest is in the attenuation length
$L_n$, not in $\gamma_n$. Therefore, it is necessary to identify
the terms that have a non-zero mean-value in the general
expression \eqref{Gen-Xi-Approx}.

\section{Fluctuating surface scattering potential}
\label{sec-Xi-tilde}

\subsection{Waveguide with uncorrelated boundaries}
\label{subsec-UB}

One can see that the mean value of the kernel in
Eq.~\eqref{Gen-Xi-Approx} is defined by the SGS-terms that are
proportional to $\sigma^2{\xi_i'}^2(x)$. In order to construct the
fluctuating part of $\Xi(k_x,k'_x)$, we introduce the {\it
square-gradient functions} $\widehat{\cal V}_{i}(x)$ for the lower
($i=\downarrow$) and upper ($i=\uparrow$) boundary, by subtracting
the mean values, $\langle{\xi_{\downarrow}'}^2(x)\rangle=
\langle{\xi_{\uparrow}'}^2(x)\rangle=\langle{\xi'}^2(x)\rangle$,
from ${\xi'_{i}}^2(x)$,
\begin{equation}\label{V-def}
\widehat{\cal V}_{i}(x)={\xi_{i}'}^2(x)
-\langle{\xi'}^2(x)\rangle\,,\qquad
\langle\widehat{\cal V}_{i}(x)\rangle=0.
\end{equation}
The functions $\widehat{\cal V}_i(x)$ play an important role in
our further consideration. In accordance with the Gaussian nature
of the surface-profile functions $\xi_{i}(x)$, there are no
correlations between $\widehat{\cal V}_i(x)$ and both $\xi_i(x)$
and $\xi_i'(x)$,
\begin{equation}\label{Corr_V_xi}
\langle\xi_i(x)\widehat{\cal V}_i(x')\rangle=0\,,\qquad
\langle\xi_i'(x)\widehat{\cal V}_i(x')\rangle=0\,,
\end{equation}
as well as between the product
$\xi_{\downarrow}'(x)\xi_{\uparrow}'(x)$ and $\widehat{\cal
V}_i(x)$, $\xi_i(x)$ or $\xi_i'(x)$.

The binary correlators for $\widehat{\cal V}_i(x)$ are given by
\begin{equation}\label{V-corr}
\langle{\widehat{\cal V}_i(x)}{\widehat{\cal V}_i(x')}\rangle=
2\langle\xi_i'(x)\xi_i'(x')\rangle^2=2{{\cal W}''}^2(x-x').
\end{equation}
Correspondingly, the Fourier transform of ${\widehat{\cal
V}_i(x)}$ reads,
\begin{equation} \label{FT-VV}
\langle V_i(k_x)V_i(k_x')\rangle=4\pi\delta(k_x+k_x')\,T(k_x)\,,
\end{equation}
due to the standard relation,
\begin{eqnarray}
V_i(k_x)&=&\int_{-\infty}^{\infty}dx\exp(-ik_xx)\widehat{\cal
V}_i(x). \label{FT-V}
\end{eqnarray}
Here we have introduced the {\it roughness-square-gradient power}
(RSGP) spectrum $T(k_x)$,
\begin{equation} \label{T-def}
T(k_x)=\int_{-\infty}^{\infty}dx\exp{(-ik_xx)}{{\cal W}''}^2(x).
\end{equation}
We also introduce the Fourier transform of the product
$\xi'_{\downarrow}(x)\xi'_{\uparrow}(x)$ and its binary correlator
,
\begin{eqnarray}
\label{FT-xi1-xi2}
&&V_{\updownarrow}(k_x)
=\int_{-\infty}^{\infty}dx\exp(-ik_xx)\xi'_{\downarrow}(x)
\xi'_{\uparrow}(x),\\[6pt]
\label{corr-FT-xi1-xi2}
&&\langle V_{\updownarrow}(k_x)V_{\updownarrow}(k_x')\rangle=
2\pi\delta(k_x+k_x')\,T(k_x),
\end{eqnarray}
as well as the Fourier transform of $\xi_i(x)$,
\begin{eqnarray}
\widetilde\xi_i(k_x)&=&\int_{-\infty}^{\infty}dx\,\exp(-ik_xx)\,
\xi_i(x).
\label{FT-xi}
\end{eqnarray}

It should be pointed out that by the integration by parts the
power spectrum of the roughness gradients $\xi_i'(x)$ can be
reduced to the RHP spectrum $W(k_x)$. However it is not possible
to do for the RSGP spectrum $T(k_x)$. This very fact reflects a
highly non-trivial role of the SGS- and GIS-terms, since they can
compete with the other terms, although they are proportional to
$\sigma^2$.

We can write now the fluctuating part of the total scattering
potential in the form,
\begin{eqnarray}
&&\widetilde \Xi(k_x,k_x')=\Xi_{1\uparrow}(k_x,k_x')
+\Xi_{1\downarrow}(k_x,k_x')
\nonumber \cr \\
&&+\Xi_{2\uparrow}(k_x,k_x')+\Xi_{2\downarrow}(k_x,k_x')
+\Xi_{2\updownarrow}(k_x,k_x'). \qquad \label{UB-Xi-tilde-def}
\end{eqnarray}
Here the first term is associated with the terms depending in
Eq.~\eqref{Gen-Xi-Approx} on $\sigma\xi_{\uparrow}(x)$ and
$\sigma\xi_{\uparrow}'(x)$, therefore, on the amplitude and
gradient scattering mechanisms, respectively,
\begin{subequations}\label{UB-Xi1u}
\begin{eqnarray}
&&\Xi_{1\uparrow}(k_x,k_x')= \nonumber \\
&&\frac{\sigma}{d} \int_{-\infty}^{\infty}dx_1
\int_0^ddz_1 \exp(-ik_xx_1)\,\frac{\sin(k_zz_1)}{k_z}
\nonumber\cr\\
&&\times\left\{2\xi_{\uparrow}(x_1)\frac{\partial^2}{\partial z_1^2}
+i(k_x+k_x')\xi_{\uparrow}'(x_1)\right.
\nonumber\cr\\
&&\left.\times\left[\frac{1}{2}+z_1\frac{\partial}{\partial z_1}
\right]\right\}\exp(ik_x'x_1)\,\frac{\sin(k_z'z_1)}{k_z'} \qquad
\label{UB-Xi1u-def}\\[6pt]
&&\Xi_{1\uparrow}(k_n,k_{n'})= \nonumber \\
&&-\sigma\{\delta_{nn'}+\cos[\pi(n-n')](1-\delta_{nn'})\}
\tilde\xi_{\uparrow}(k_n-k_{n'}). \nonumber \\
\label{UB-Xi1u-exp}
\end{eqnarray}
\end{subequations}
The second term in Eq.~\eqref{UB-Xi-tilde-def} corresponding to
the lower boundary, is of the same form,
\begin{subequations}\label{UB-Xi1d}
\begin{eqnarray}
&&\Xi_{1\downarrow}(k_x,k_x')= \nonumber \\
&&-\frac{\sigma}{d} \int_{-\infty}^{\infty}dx_1 \int_0^ddz_1
\exp(-ik_xx_1)\,\frac{\sin(k_zz_1)}{k_z}
\nonumber\cr\\
&&\times\left\{2\xi_{\downarrow}(x_1)\frac{\partial^2}{\partial z_1^2}
+i(k_x+k_x')\xi_{\downarrow}'(x_1)\right. \nonumber\cr\\
&&\left.\times\left[\frac{1}{2}+\left(z_1-d\right)
\frac{\partial}{\partial z_1}\right]\right\}
\exp(ik_x'x_1)\,\frac{\sin(k_z'z_1)}{k_z'} \qquad\quad
\label{UB-Xi1d-def}\\[6pt]
&&\Xi_{1\downarrow}(k_n, k_{n'})=\sigma\,
\tilde\xi_{\downarrow}(k_n-k_{n'}). \label{UB-Xi1d-exp}
\end{eqnarray}
\end{subequations}
Next two terms in Eq.~\eqref{UB-Xi-tilde-def} are the SGS-terms,
with ${\xi_{i}'}^2(x)$ being replaced by $\hat{\cal V}_{i}(x)$,
\begin{subequations}\label{UB-Xi2u}
\begin{eqnarray}
&&\Xi_{2\uparrow}(k_x,k_x')= \nonumber \\
&&-\, \frac{\sigma^2}{d^2} \int_{-\infty}^{\infty}dx_1\int_0^ddz_1
\exp(-ik_xx_1)\,\frac{\sin(k_zz_1)}{k_z}
\nonumber\cr\\
&&\times \hat{\cal V}_{\uparrow}(x_1)\left[\frac{1}{4}+
2z_1\,\frac{\partial}{\partial z_1}+
z_1^2\frac{\partial^2}{\partial z_1^2}\right]
\nonumber\cr\\
&&\times\exp(ik_x'x_1)\,\frac{\sin(k_z'z_1)}{k_z'}
\label{UB-Xi2u-def}\\[6pt]
&&\Xi_{2\uparrow}(k_n,k_{n'})= \nonumber \\
&&\frac{\sigma^2d}{2} \bigg\{\left[\frac{1}{3}+ \frac{1}{(2\pi
n)^2}\right]\delta_{nn'}
+\frac{4}{\pi^2}\,\frac{n^2+n'^2}{(n^2-n'^2)^2} \nonumber\cr \\
&&\times \cos[\pi(n-n')]\, (1-\delta_{nn'}) \bigg\}\,
V_{\uparrow}(k_n-k_{n'}). \qquad \label{UB-Xi2u-exp}
\end{eqnarray}
\end{subequations}
\begin{subequations}\label{UB-Xi2d}
\begin{eqnarray}
&&\Xi_{2\downarrow}(k_x,k_x')= \nonumber \\
&&-\, \frac{\sigma^2}{d^2} \int_{-\infty}^{\infty}dx_1\int_0^ddz_1
\exp(-ik_xx_1)\,\frac{\sin(k_zz_1)}{k_z}
\nonumber\cr\\
&&\times \hat{\cal V}_{\downarrow}(x_1)\left[\frac{1}{4}+
2\left(z_1-d\right)\,\frac{\partial}{\partial z_1}+
\left(z_1-d\right)^2\frac{\partial^2}{\partial z_1^2}\right]
\nonumber\cr\\
&&\times\exp(ik_x'x_1)\,\frac{\sin(k_z'z_1)}{k_z'}
\label{UB-Xi2d-def}\\[6pt]
&&\Xi_{2\downarrow}(k_n,k_{n'})= \nonumber \\
&&\frac{\sigma^2d}{2}\bigg\{\left[\frac{1}{3}+
\frac{1}{(2\pi n)^2}\right]\delta_{nn'} \nonumber\cr\\
&&+\frac{4}{\pi^2}\,\frac{n^2+n'^2}{(n^2-n'^2)^2}\, (1-\delta_{nn'})
\bigg\}\,V_{\downarrow}(k_n-k_{n'}).\qquad\quad \label{UB-Xi2d-exp}
\end{eqnarray}
\end{subequations}
Finally, the last term in Eq.~\eqref{UB-Xi-tilde-def} is due to
the GIS terms in Eq.~\eqref{Gen-Xi-Approx},
\begin{subequations}\label{UB-Xi2lu}
\begin{eqnarray}
&&\Xi_{2\updownarrow}(k_x,k_x')= \nonumber \\
&&\frac{\sigma^2}{d^2} \int_{-\infty}^{\infty}dx_1\int_0^ddz_1
\exp(-ik_xx_1)\,\frac{\sin(k_zz_1)}{k_z}
\nonumber\cr\\
&&\times \xi_{\downarrow}'(x_1)\xi_{\uparrow}'(x_1) \left[\frac{1}{2}+
2(2z_1-d)\,\frac{\partial}{\partial z_1}+
2z_1(z_1-d)\frac{\partial^2}{\partial z_1^2}\right]
\nonumber\cr\\
&&\times\exp(ik_x'x_1)\,\frac{\sin(k_z'z_1)}{k_z'}
\label{UB-Xi2lu-def}\\[6pt]
&&\Xi_{2\updownarrow}(k_n, k_{n'})= \nonumber \\
&&\frac{\sigma^2d}{2}\bigg\{\left[\frac{1}{3}- \frac{1}{2(\pi
n)^2}\right]\delta_{nn'}-\frac{8}{\pi^2}\,
\frac{n^2+n'^2}{(n^2-n'^2)^2} \nonumber\cr\\
&&\times\cos^2[\pi(n-n')/2]\,(1-\delta_{nn'}) \bigg\}\,
V_{\updownarrow}(k_n-k_{n'}). \label{UB-Xi2lu-exp}
\end{eqnarray}
\end{subequations}

The kernel $\widetilde\Xi(k_x,k_x')$ written as the sum
\eqref{UB-Xi-tilde-def} of specially designed terms, has a very
convenient form. First, all the terms are chosen to have zero
average. Second, since the functions $\xi_i(x)$,
$\xi_{\downarrow}(x)\xi_{\uparrow}(x)$ and $\widehat{\cal
V}_{i}(x)$ do not correlate with each other, their Fourier
transforms are also non-correlated:
\begin{subequations}\label{FT-Corr}
\begin{eqnarray}
\label{FT-Corr-Xi-V}
&&\langle\tilde\xi_{i}(k_x) V_{i}(k_x')\rangle=0. \\
&&\langle\tilde\xi_{i}(k_x) V_{\updownarrow}(k_x')\rangle=0. \\
&&\langle V_{i}(k_x') V_{\updownarrow}(k_x) \rangle=0.
\end{eqnarray}
\end{subequations}
Therefore, all the terms in Eq.~\eqref{UB-Xi-tilde-def}
determining the potential, are also non-correlated. Their
autocorrelators are,
\begin{subequations}\label{UB-Xi1-corr-Q1}
\begin{align} \label{UB-Xi1-corr}
&\langle\Xi_{1i}(k_n,k_{n'})\Xi_{1i}(k_{n'},k_{n''})\rangle=
2\pi\delta(k_n-k_{n''})Q_1(k_n,k_{n'}),\raisetag{-2pt} \\[6pt]
\label{UB-Q1} &Q_1(k_n,k_{n'})=\sigma^2W(k_n-k_{n'}),
\end{align}
\end{subequations}
\begin{subequations}\label{UB-Xi2-corr-Q2}
\begin{align}\label{UB-Xi2-corr}
&\langle\Xi_{2i}(k_n,k_{n'})\Xi_{2i}(k_{n'},k_{n''})\rangle=
2\pi\delta(k_n-k_{n''})Q_2(k_n,k_{n'}),\raisetag{-2pt} \\[6pt]
\label{UB-Q2} &Q_2(k_n,k_{n'})=\frac{\sigma^4d^2}{2}T(k_n-k_{n'})
\bigg\{\left[\frac{1}{3}+\frac{1}{(2\pi n)^2}\right]^2\delta_{nn'} \notag \\
&+\frac{16}{\pi^4}\,\frac{(n^2+n'^2)^2}{(n^2-n'^2)^4}\,
(1-\delta_{nn'})\bigg\},
\end{align}
\end{subequations}
\begin{subequations}\label{UB-Xi2lu-corr-Q2lu}
\begin{align}\label{UB-Xi2lu-corr}
&\langle\Xi_{2\updownarrow}(k_n,k_{n'})
\Xi_{2\updownarrow}(k_{n'},k_{n''})\rangle
=2\pi\delta(k_n-k_{n''})Q_{2\updownarrow}(k_n,k_{n'})
\raisetag{-2pt} \\[6pt] \label{UB-Q2lu}
&Q_{2\updownarrow}(k_n,k_{n'})=\frac{\sigma^4d^2}{4}T(k_n-k_{n'})
\left\{\left[\frac{1}{3}-
\frac{1}{2(\pi n)^2}\right]^2\delta_{nn'}\right. \nonumber \\
&\left.+\frac{64}{\pi^4}\,\frac{(n^2+n'^2)^2}{(n^2-n'^2)^4}\,
\cos^4[\pi(n-n')/2]\,(1-\delta_{nn'})\right\}. \raisetag{-5pt}
\end{align}
\end{subequations}
Note that in spite of the absence of correlations between the
upper and lower surface profiles, there are the correlations due
to their derivatives since $\langle\Xi_{2\updownarrow}(k_n,k_{n'})
\Xi_{2\updownarrow}(k_{n'},k_{n''})\rangle \neq 0$ (see
Eq.~\eqref{UB-Xi2lu-corr}).

As a result, the correlator of the fluctuating scattering
potential \eqref{UB-Xi-tilde-def} in the $k_x$-representation
reads,
\begin{subequations}\label{UB-tilde-Xi-corr}
\begin{align}
\label{UB-Xi-Q}
&\langle\widetilde\Xi(k_n,k_{n'})\widetilde\Xi(k_{n'},k_{n''})\rangle=
2\pi\delta(k_n-k_{n''})Q(k_n,k_{n'}), \\[6pt]
\label{UB-Q} &Q(k_n,k_{n'})=2Q_1(k_n,k_{n'})+2Q_2(k_n,k_{n'})
+Q_{2\updownarrow}(k_n,k_{n'})\,. \raisetag{-4pt}
\end{align}
\end{subequations}
The factor $2$ in the first and second terms arises due to
independent contributions, $Q_1(k_n,k_{n'})$ and
$Q_2(k_n,k_{n'})$, from the lower and upper boundaries.

\subsection{Waveguide with antisymmetric rough boundaries}
\label{subsec-CCB}

In this configuration both boundaries of the waveguide are
identical, see Eq.~\eqref{CCBcase}, therefore, antisymmetric with
respect to the central line $z=d/2$. For this specific case the
width of the waveguide (the distance between upper and lower
boundaries) remains constant along the $x-$direction. Instead of
Eq.~~\eqref{Gen-Xi-Approx} we have to start now with the exact
expression (\eqref{Gen-Xi-exactF}) by substituting the relation
(\eqref{CCBcase}). As a result, one gets the relatively simple
expression for the scattering potential,
\begin{equation}\label{CCB-Xi}
\begin{split}
&\Xi(k_x,k'_x)=\int_{-\infty}^{\infty}dx_1\int_0^ddz_1\,
\exp\left(-ik_xx_1\right) \frac{\sin(k_zz_1)}{k_z} \\
&\times \Biggl(\sigma\, [i(k_x+k_x')]\,
\xi'(x_1)\,\frac{\partial}{\partial z_1} \\
&-\sigma^2\,{\xi'}^2(x_1)\frac{\partial^2}{{\partial z_1}^2}\Bigg)
\exp\left(ik'_xx_1\right)\frac{\sin(k'_zz_1)}{k'_z}
\end{split}
\end{equation}
One can realize that all terms proportional to $\sigma^2$ can be
written as one SGS-term with a non-zero mean value. By making use
of the functions defined by Eqs.~\eqref{V-def}, \eqref{FT-V} and
\eqref{FT-xi} with $i=\uparrow$, in what follows we omit the
subscript $i$.

Now the fluctuating part of the scattering potential is,
\begin{equation}\label{CCB-Xi-tilde-def}
\widetilde \Xi(k_x,k_x')=\Xi_{1}(k_x,k_x')+\Xi_{2}(k_x,k_x').
\end{equation}
The first summand related to the gradient scattering terms in
Eq.~\eqref{CCB-Xi} has the form,
\begin{subequations}\label{CCB-Xi1}
\begin{eqnarray}
&&\Xi_1(k_x,k_x')=\sigma\int_{-\infty}^{\infty}dx_1\int_0^ddz_1
\exp(-ik_xx_1)\,\frac{\sin(k_zz_1)}{k_z}
\nonumber\cr\\
&&\times [i(k_x+k_x')]\,\xi'(x_1)\,\frac{\partial}{\partial z_1}
\exp(ik_x'x_1)\,\frac{\sin(k_z'z_1)}{k_z'}
\label{CCB-Xi1-def}\\[6pt]
&&\Xi_1(k_n,k_{n'})=2\sigma\sin^2[\pi(n-n')/2]\,
\tilde\xi(k_n-k_{n'}). \label{CCB-Xi1-exp}
\end{eqnarray}
\end{subequations}
The second summand is due to the SGS-term, however, with the
function ${\hat{\cal V}(x_1)}$ instead of ${\xi'}^2(x_1)$. With
the use of the Fourier transform (\ref{FT-V}) for the operator
${\hat{\cal V}(x)}$, one gets,
\begin{subequations}\label{CCB-Xi2}
\begin{eqnarray}
&&\Xi_2(k_x,k_x')=-\int_{-\infty}^{\infty}dx_1\int_0^ddz_1
\exp(-ik_xx_1)\,\frac{\sin(k_zz_1)}{k_z}
\nonumber\cr\\
&&\times\sigma^2\hat{\cal V}(x_1)\,\frac{\partial^2}{\partial
z_1^2}
\exp(ik_x'x_1)\,\frac{\sin(k_z'z_1)}{k_z'}
\label{CCB-Xi2-def}\\[6pt]
&&\Xi_2(k_n,k_{n'})=\frac{\sigma^2d}{2}\,\delta_{nn'}\,
V(k_n-k_{n'}). \label{CCB-Xi2-exp}
\end{eqnarray}
\end{subequations}
Note that the expressions (\ref{CCB-Xi1}) and (\ref{CCB-Xi2}) for
the scattering potential are still {\it exact}. In contrast with
other configurations discussed below, the expansion in
$\sigma/d$-smallness is not required because of the constant width
of the waveguide.

The correlation properties of the scattering potential are
described by the following formulas. Due to the condition
(\ref{FT-Corr-Xi-V}), the terms $\Xi_1(k_x,k_x')$ and
$\Xi_2(k_x,k_x')$ are non-correlated,
\begin{equation}\label{CCB-Xi1(k)-Xi2(k)}
\langle\Xi_1(k_x,k_x')\Xi_2(k_x',k_x'')\rangle=0.
\end{equation}
On the other hand, the autocorrelators are non-zero,
\begin{subequations}\label{CCB-Xi1-corr-Q1}
\begin{eqnarray}
&&\langle\Xi_1(k_n,k_{n'})\Xi_1(k_{n'},k_{n''})\rangle=
2\pi\delta(k_n-k_{n''})Q_1(k_n,k_{n'}),\nonumber\\\label{CCB-Xi1-corr}\\
&&Q_1(k_n,k_{n'})=4\sigma^2 W(k_n-k_{n'})\sin^4[\pi(n-n')/2].
\nonumber\\\label{CCB-Q1}
\end{eqnarray}
\end{subequations}
\begin{subequations}\label{CCB-Xi2-corr-Q2}
\begin{eqnarray}
&&\langle\Xi_2(k_n,k_{n'})\Xi_2(k_{n'},k_{n''})\rangle=
2\pi\delta(k_n-k_{n''})Q_2(k_n,k_{n'}),\nonumber\\\label{CCB-Xi2-corr}\\
&&Q_2(k_n,k_{n'})=(\sigma^4d^2/2)\,T(k_n-k_{n'})\,\delta_{nn'}.
\nonumber\\\label{CCB-Q2}
\end{eqnarray}
\end{subequations}
Finally, the correlator of the fluctuating part of the total
scattering potential $\widetilde \Xi(k_x,k_x')$ has the following
form,
\begin{subequations}\label{CCB-Xi-tilde-corr-Q}
\begin{eqnarray}
&&\langle\widetilde\Xi(k_n,k_{n'})\widetilde\Xi(k_{n'},k_{n''})\rangle=
2\pi\delta(k_n-k_{n''})Q(k_n,k_{n'}),\nonumber\\\label{CCB-Xi-tilde-corr}\\
&&Q(k_n,k_{n'})=Q_1(k_n,k_{n'})+Q_2(k_n,k_{n'}).\label{CCB-Q}
\end{eqnarray}
\end{subequations}

\subsection{Waveguide with symmetric rough
boundaries} \label{subsec-SB}
Let us now consider the waveguide with symmetric boundaries, see
Eq.~(\ref{SBcase}), therefore, with the width defined by
\begin{equation}\label{SB-d(x)}
w(x)=d+2\sigma\xi(x), \qquad \langle w(x)\rangle=d.
\end{equation}
By substituting the symmetric profile into
Eq.~\eqref{Gen-Xi-Approx}, we obtain for the surface scattering
potential,
\begin{eqnarray}\label{SB-Xi-Approx}
&&\Xi(k_x,k'_x)\approx\int_{-\infty}^{\infty}dx_1\int_0^ddz_1\,
\exp\left(-ik_xx_1\right)\frac{\sin(k_zz_1)}{k_z}
\nonumber\cr\\
&&\times \left\{\frac{4\sigma\xi(x_1)}{d}
\frac{\partial^2}{\partial z_1^2}\right.
\nonumber\cr\\
&&+i(k_x+k_x')\frac{\sigma\xi'(x_1)}{d}\left[1+\left(2z_1-d\right)
\frac{\partial}{\partial z_1}\right]
\nonumber\cr\\
&&-\left.\frac{\sigma^2{\xi'}^2(x_1)}{d^2}\left[1+4\left(2z_1-d\right)\,
\frac{\partial}{\partial z_1}+
\left(2z_1-d\right)^2\frac{\partial^2}{\partial z_1^2}\right]\right\}
\nonumber\cr\\
&&\times\exp\left(ik'_xx_1\right)\frac{\sin(k'_zz_1)}{k'_z}\,.
\end{eqnarray}
As for the previous configuration, see Eq.~\eqref{CCB-Xi}, the
SGS-terms corresponding to the lower boundary and the GIS-terms
are written in terms of $\xi(x)=\xi_{\uparrow}(x)$.

The potential in Eq.~(\ref{SB-Xi-Approx}) can be written as the
sum of two terms related to substantially different scattering
mechanisms,
\begin{equation}\label{SB-Xi-tilde-def}
\widetilde \Xi(k_x,k_x')=\Xi_{1}(k_x,k_x')+\Xi_{2}(k_x,k_x').
\end{equation}
Here the first summand is related to the terms proportional to
$\sigma$ in Eq.~\eqref{SB-Xi-Approx} (the amplitude and gradient
scattering terms),
\begin{subequations}\label{SB-Xi1}
\begin{eqnarray}
&&\Xi_1(k_x,k_x')=\int_{-\infty}^{\infty}dx_1\int_0^ddz_1
\exp(-ik_xx_1)\,\frac{\sin(k_zz_1)}{k_z}
\nonumber\cr\\
&&\times\left\{\frac{4\sigma\xi(x_1)}{d}\frac{\partial^2}{\partial
z_1^2}+i(k_x+k_x')\frac{\sigma\xi'(x_1)}{d}\right.\nonumber\cr\\
&&\left.\times\left[1+\left(2z_1-d\right)
\frac{\partial}{\partial z_1}\right]\right\}
\exp(ik_x'x_1)\,\frac{\sin(k_z'z_1)}{k_z'}
\label{SB-Xi1-def}\\[6pt]
&&\Xi_1(k_n,k_{n'})=-2\sigma\{\delta_{nn'}+\cos^2[\pi(n-n')/2]
\nonumber\\
&&\times(1-\delta_{nn'})\}\,\tilde\xi(k_n-k_{n'}).\nonumber\\
\label{SB-Xi1-exp}
\end{eqnarray}
\end{subequations}
The second summand is associated with the fluctuating part of the
SGS-terms in Eq.~(\ref{SB-Xi-Approx}), and consequently, with the
function ${\xi'}^2(x)$ being replaced by $\hat{\cal V}(x)$,
\begin{subequations}\label{SB-Xi2}
\begin{eqnarray}
&&\Xi_2(k_x,k_x')=-\int_{-\infty}^{\infty}dx_1\int_0^ddz_1
\exp(-ik_xx_1)\,\frac{\sin(k_zz_1)}{k_z}
\nonumber\cr\\
&&\times\frac{\sigma^2\hat{\cal V}(x_1)}{d^2}\left[1+
4\left(2z_1-d\right)\,\frac{\partial}{\partial z_1}+
\left(2z_1-d\right)^2\frac{\partial^2}{\partial z_1^2}\right]
\nonumber\cr\\
&&\times\exp(ik_x'x_1)\,\frac{\sin(k_z'z_1)}{k_z'}
\label{SB-Xi2-def}\\[6pt]
&&\Xi_2(k_n,k_{n'})=\frac{\sigma^2d}{2} \bigg\{\left[\frac{1}{3}+
\frac{1}{(\pi n)^2}\right]\delta_{nn'}+\frac{16}{\pi^2}\,\nonumber\cr\\
&&\times\frac{n^2+n'^2}{(n^2-n'^2)^2}
\cos^2[\pi(n-n')/2](1-\delta_{nn'})\bigg\}V(k_n-k_{n'}).\nonumber\\
\label{SB-Xi2-exp}
\end{eqnarray}
\end{subequations}

The correlation properties of the introduced surface scattering
potentials $\Xi_1(k_n,k_{n'})$ and $\Xi_2(k_n,k_{n'})$ are
described by the correlator,
\begin{subequations}\label{SB-Xi1-corr-Q1}
\begin{eqnarray}
&&\langle\Xi_1(k_n,k_{n'})\Xi_1(k_{n'},k_{n''})\rangle=
2\pi\delta(k_n-k_{n''})Q_1(k_n,k_{n'}),\nonumber\\\label{SB-Xi1-corr}\\
&&Q_1(k_n,k_{n'})=4\sigma^2 W(k_n-k_{n'})\nonumber\\
&&\times \{\delta_{nn'}+\cos^4[\pi(n-n')/2](1-\delta_{nn'})\}.
\label{SB-Q1}
\end{eqnarray}
\end{subequations}
\begin{subequations}\label{SB-Xi2-corr-Q2}
\begin{eqnarray}
&&\langle\Xi_2(k_n,k_{n'})\Xi_2(k_{n'},k_{n''})\rangle=
2\pi\delta(k_n-k_{n''})Q_2(k_n,k_{n'}),\nonumber \\\label{SB-Xi2-corr}\\
&&Q_2(k_n,k_{n'})=\frac{\sigma^4d^2}{2}T(k_n-k_x')\,
\bigg\{\left[\frac{1}{3}+\frac{1}{(\pi n)^2}\right]^2
\delta_{nn'}\nonumber\\
&&+\frac{256}{\pi^4}\,\frac{(n^2+n'^2)^2}{(n^2-n'^2)^4}
\cos^4[\pi(n-n')/2](1-\delta_{nn'})\bigg\}.\nonumber\\
\label{SB-Q2}
\end{eqnarray}
\end{subequations}
Thus, we arrive at the expression,
\begin{subequations}\label{SB-Xi-tilde-corr-Q}
\begin{eqnarray}
&&\langle\widetilde\Xi(k_n,k_{n'})\widetilde\Xi(k_{n'},k_{n''})\rangle=
2\pi\delta(k_n-k_{n''})Q(k_n,k_{n'}),\nonumber\\\label{SB-Xi-tilde-corr}\\
&&Q(k_n,k_{n'})=Q_1(k_n,k_{n'})+Q_2(k_n,k_{n'}).\label{SB-Q}
\end{eqnarray}
\end{subequations}

\section{Average Green's function and attenuation length}
\label{sec-averageGF}

Now we replace the problem for the random Green's function
$g(k_x,k_x')$ with the problem for the Green's function $\langle
g(k_x,k_x')\rangle$ averaged over the surface disorder. To perform
the averaging of Eq.~(\ref{g-DE}) with $\Xi(k_x,k_x')$, given for
the uncorrelated, antisymmetric and symmetric boundary
configurations by Eq.~\eqref{UB-Xi-tilde-def},
\eqref{CCB-Xi-tilde-def} and \eqref{SB-Xi-tilde-def},
respectively, and obtain $\langle g(k_x,k_x')\rangle$, we can
apply one of the standard and well known perturbative methods. For
example, it can be the diagrammatic approach developed for surface
disordered systems~\cite{BFb79}, as well as the technique
developed in Ref.~\onlinecite{McGM84}. Both of the methods take
adequately into account the multiple scattering from the
corrugated boundaries, and allow one to develop the consistent
perturbative approach with respect to the scattering potential.
After quite cumbersome calculations (see Ref.~\cite{IzMkRn-PRB-06}
for details), we obtain the average Green's function in the
normal-mode representation,
\begin{eqnarray}
&&\langle{\cal G}(x,x';z,z')\rangle=\sum_{n=1}^{N_d}
\sin\left(\frac{\pi nz}{d}\right) \sin\left(\frac{\pi
nz'}{d}\right)
\nonumber\cr\\
&&\times\frac{\exp(ik_n|x-x'|)}{ik_nd}\,
\exp\left(-\frac{|x-x'|}{2L_n}\right).
\label{<GF>}
\end{eqnarray}
The quantity $L_n$ is the wave attenuation length or electron
total mean-free-path of the $n$th conducting mode. This quantity
is determined by the imaginary part of the self-energy and
describes the scattering from the $n$th mode into other
propagating modes,
\begin{eqnarray}
\frac{1}{L_n}&=&\frac{(\pi n/d)^2}{k_nd}\,\sum_{n'=1}^{N_d}
\frac{(\pi n'/d)^2}{k_{n'}d}
\nonumber\\[6pt]
&&\times\left[Q(k_n,-k_{n'})+Q(k_n,+k_{n'})\right].
\label{Ln-Q}
\end{eqnarray}
Here the explicit expressions for $Q(k_n,-k_{n'})$ and
$Q(k_n,+k_{n'})$ are given by Eqs.~\eqref{UB-Q}, \eqref{CCB-Q} and
\eqref{SB-Q} for the waveguide with uncorrelated, antisymmetric
and symmetric boundaries.

In deriving Eqs.~(\ref{<GF>}) and (\ref{Ln-Q}) we have made the
following simplifications. First, the self-energy in the
Dyson-type equation for the average Green's function was obtained
within the second-order approximation in the perturbation
potential. In terms of the diagrammatic technique this is similar
to the ``simple vortex" or, the same, Bourret
approximation~\cite{Bourret62} which contains the binary
correlator $Q(k_x,k_x')$ of the surface scattering potential and
the unperturbed pole factor $g_0(q_x)$. Second, in order to
extract the inverse attenuation length from the self-energy, we
substituted the lengthwise wave number $k_x$ by its unperturbed
value $k_n$. The justification of this substitution, as well as
the conditions of applicability resulting from the above
simplifications, were discussed in Ref.~\cite{IzMkRn-PRB-06}.
Specifically, it was shown that the domain of applicability is
restricted by two independent criteria of the {\it weak surface
scattering},
\begin{equation}\label{WS-Ln}
\Lambda_n=k_nd/(\pi n/d) \ll2L_n, \qquad R_c\ll2L_n.
\end{equation}
Here $\Lambda_n$ is the distance between two successive
reflections of the $n$th mode from one boundary to the opposite
one. The conditions in \eqref{WS-Ln} imply that the wave is weakly
attenuated on both the correlation length $R_c$ and the cycle
length $\Lambda_n$. They implicitly include the requirement that
$\sigma\ll d$ for the surface corrugations be small in height,
used in the Sec.~\ref{sec-Xi-tilde} when deriving the explicit
form (\ref{Gen-Xi-Approx}) for the surface scattering potential
$\Xi(k_x,k_x')$.

\section{Attenuation Length Analysis}
\label{sec-Ln-An}

In this section we discuss the attenuation length $L_n$ for three
configurations of the waveguide. In the first part of the
subsections we give explicit expressions for the $1/L_n$ for any
form of the binary correlator ${\cal W}(x)$. In the second part,
we assume that the random surface profile $\xi(x)$ has the
Gaussian binary correlator,
\begin{equation}\label{corr-Gauss}
{\cal W}(x)=\exp(-x^2/2R_c^2).
\end{equation}
Then the RHP spectrum (\ref{FT-W}) is given by
\begin{equation}\label{WFT-Gaus}
W(k_x)=\sqrt{2\pi}\,R_c\,\exp(-k_x^2R_c^2/2).
\end{equation}
The RSGP spectrum $T(k_x)$ defined by Eq.~(\ref{T-def})), can be
written as
\begin{eqnarray}
T(k_x)
&=&\frac{\sqrt{\pi}}{16R_c^3}\left[(k_xR_c)^4-4(k_xR_c)^2+12\right]
\nonumber\\[6pt]
&\times&\exp\left[-(k_xR_c)^2/4\right]\,.
\label{T-Gaus-expl}
\end{eqnarray}

In what follows, we express the inverse attenuation length in the
form of the dimensionless quantity $\Lambda_n/2L_n$. Since $L_n$
depends on as many as four dimensionless parameters $(k\sigma)^2$,
$kR_c$, $kd/\pi$, and $n$, a complete analysis appears to be quite
complicated. For this reason, below we restrict ourselves to an
analysis of the interplay between different scattering mechanisms,
as a function of the dimensionless correlation length $kR_c$ for
different values of $(k\sigma)^2$. We concentrate our attention on
a multimode waveguide with a large number of propagating modes,
$N_d\approx kd/\pi\gg 1$. Our main interest is in two situations,
namely, in a small-scale roughness of the ``white noise'' kind for
which $kR_c\ll 1$, and in large-scale random corrugations with
$kR_c \gg 1$.

\subsection{Waveguide with uncorrelated boundaries}
\label{subsec-UB-L}

In view of Eq.~(\ref{UB-Q}), the general expression (\ref{Ln-Q})
for the inverse attenuation length shows that for a waveguide with
uncorrelated boundaries this quantity consists of three terms,
\begin{equation}\label{UB-Ln-sum}
\frac{1}{L_n}=\frac{1}{L^{(1)}_n}+\frac{1}{L^{(2)}_n}
+\frac{1}{L^{(2)\updownarrow}_n}.
\end{equation}
These terms are originated from different mechanisms of the
surface scattering. The first attenuation length $L^{(1)}_n$ is
related to the RHP spectrum $W(k_x)$ through the expression for
$2Q_1(k_x,k'_x)$. In accordance with Eq.~(\ref{UB-Q1}) it is given
by
\begin{eqnarray}\label{UB-Ln1-def}
\frac{1}{L^{(1)}_n}&=&
2\sigma^2\,\frac{(\pi n/d)^2}{k_nd}\,\sum_{n'=1}^{N_d}
\frac{(\pi n'/d)^2}{k_{n'}d}\nonumber\\[6pt]
&&\times\left[ W(k_n+k_{n'})+W(k_n-k_{n'})\right].\qquad\qquad
\end{eqnarray}
Here the term $1/L^{(1)}_{n,n}$ corresponding to $n'=n$, is
related to the amplitude scattering while the terms
$1/L^{(1)}_{n,n'\neq n}$ result from the gradient scattering.
These two mechanisms of surface scattering are due to the
corresponding terms in the expression for $\Xi_{1 i}(k_x,k_x')$,
see Eqs.~\eqref{UB-Xi1u-def} and \eqref{UB-Xi1d-def}. The first
one is related to the term depending on the amplitude of the
roughness profile $\sigma\xi_i(x)$, and the second one is due to
the terms depending on the roughness gradient $\sigma\xi_i'(x)$.
The expression (\ref{UB-Ln1-def}) exactly coincides with that
previously obtained by various methods (see, e.g.,
Ref.~\onlinecite{BFb79}).

The second attenuation length $L^{(2)}_n$ related to the RSGP
spectrum through $2Q_2(k_x,k_x')$, is associated solely with the
SGS mechanism due to the operator $\widehat{\cal V}_i(x)$, see
Eqs.~\eqref{UB-Xi2u-def} and \eqref{UB-Xi2d-def}. In accordance
with Eq.~(\ref{UB-Q2}), it is described by
\begin{equation}\label{UB-Ln2}
\frac{1}{L^{(2)}_n}=\sum_{n'=1}^{N_d}\frac{1}{L^{(2)}_{n,n'}}.
\end{equation}
Here the terms in the sum are
\begin{equation}\label{UB-Lnn2}
\frac{1}{L^{(2)}_{n,n}}=\sigma^4\,\frac{(\pi
n/d)^4}{k_n^2}\,\left[\frac{1}{3}+\frac{1}{(2\pi n)^2}\right]^2
\left[T(2k_n)+T(0)\right]
\end{equation}
and
\begin{equation}\label{UB-Lnn'2}
\begin{split}
\frac{1}{L^{(2)}_{n,n'\neq n}}=&\sigma^4\,\frac{16}{\pi^4}\frac{(\pi
n/d)^2}{k_n}\,\frac{(\pi n'/d)^2}{k_{n'}}\,
\frac{(n^2+n'^2)^2}{(n^2-n'^2)^4}\\[6pt]
&\times\left[T(k_n+k_{n'})+T(k_n-k_{n'})\right].
\end{split}
\end{equation}

The third term $L^{(2)\updownarrow}_n$ is also related to the RSGP
spectrum through $Q_{2\updownarrow}(k_x,k_x')$ due to the product
$\xi_{\downarrow}'(x)\xi_{\uparrow}'(x)$, see
Eq.~(\ref{UB-Xi2lu-def}). In accordance with Eqs.~(\ref{UB-Q2lu}),
it is described by
\begin{equation}\label{UB-Ln2lu}
\frac{1}{L^{(2)\updownarrow}_n}=
\sum_{n'=1}^{N_d}\frac{1}{L^{(2)\updownarrow}_{n,n'}},
\end{equation}
where
\begin{equation}\label{UB-Lnn2lu}
\begin{split}
\frac{1}{L^{(2)\updownarrow}_{n,n}}=&\frac{\sigma^4}{4}\,\frac{(\pi
n/d)^4}{k_n^2}\,\left[\frac{1}{3}-\frac{1}{2(\pi n)^2}\right]^2 \\[6pt]
&\times\left[T(2k_n)+T(0)\right].
\end{split}
\end{equation}
and
\begin{equation}\label{UB-Lnn'2lu}
\begin{split}
&\frac{1}{L^{(2)\updownarrow}_{n,n'\neq n}}=\sigma^4\,\frac{16}{\pi^4}\,
\frac{(\pi n/d)^2}{k_n}\,\frac{(\pi n'/d)^2}{k_{n'}}\,
\frac{(n^2+n'^2)^2}{(n^2-n'^2)^4}\\[6pt]
&\times\left[T(k_n+k_{n'})+T(k_n-k_{n'})\right] \cos^4[\pi(n-n')/2].
\end{split}
\end{equation}

Analogously, the terms in Eq.~\eqref{UB-Ln-sum} can be classified
as partial terms describing the intramode ($1/L^{(1)}_{n,n}$,
$1/L^{(2)}_{n,n}$ and $1/L^{(2)\updownarrow}_{n,n}$), and
intermode  ($1/L^{(1)}_{n,n'\neq n}$, $1/L^{(2)}_{n,n'\neq n}$ and
$1/L^{(2)\updownarrow}_{n,n'\neq n}$) scattering. According to the
discussion of Eq.~\eqref{UB-Ln1-def}, the amplitude and gradient
terms result in the intramode and intermode scattering,
respectively. As for the square-gradient terms, they lead to both
types of scattering.

\subsubsection{Small-scale roughness}
\label{subsubsec-UB-Ln-SSR}

To continue our analysis, let us start with the widely used case
of {\it a small-scale boundary perturbation}, $kR_c\ll 1$. In this
case the surface roughness can be regarded as a delta-correlated
random process with the correlator ${\cal W}(x-x')\approx
W(0)\delta(x-x')$ and constant power spectrum $W(k_x)\approx
W(0)\sim R_c$. Taking into account the evident relationship
$k\Lambda_n\gtrsim 1$, one can get the following inequalities
specifying this case,
\begin{equation}\label{SSR-def}
kR_c\ll 1\lesssim k\Lambda_n.
\end{equation}
It is necessary to underline that in the regime of small-scale
roughness (\ref{SSR-def}) the second of the weak-scattering
conditions in Eq.(\ref{WS-Ln}) is not so restrictive as the first
one, and directly stems from it, $R_c\ll\Lambda_n\ll2L_n$.

In Eqs.~(\ref{UB-Ln1-def}), (\ref{UB-Lnn2}), (\ref{UB-Lnn'2}),
(\ref{UB-Lnn2lu}) and (\ref{UB-Lnn'2lu}) for the attenuation lengths
the argument of the correlators $W(k_x)$ and $T(k_x)$ turns out to
be much less than the scale of their decrease $R_c^{-1}$ under the
conditions (\ref{SSR-def}). Thus, for any term in the summation over
$n'$ the argument can be taken as zero.

Therefore, the first attenuation length is determined as follows,
\begin{subequations}\label{UB-Ln1-SSR}
\begin{eqnarray}
&&\frac{\Lambda_n}{2L^{(1)}_n}\approx
2(k\sigma)^2\frac{n}{kd/\pi}\frac{W(0)}{k}\sum_{n'=1}^{N_d}
\frac{(\pi n'/d)^2}{k_{n'}d}\qquad
\label{UB-Ln1-SSR-Sum}\\[6pt]
&&\approx (k\sigma)^2\frac{n}{kd/\pi}\,\frac{kW(0)}{2}
\label{UB-Ln1-SSR-W}
\end{eqnarray}
\end{subequations}
The approximation is made here for a large number of the
conducting modes, with the change of the summation over $n'$ by
integration. For a further estimate, one can take into account the
formula
\begin{equation}\label{W(0)}
W(0)=\int_{-\infty}^{\infty}dx{\cal W}(x)
=2R_c\int_{0}^{\infty}d\rho{\cal W}(R_c\rho),
\end{equation}
that directly follows from the definition (\ref{FT-W}) for the
Fourier transform $W(k_x)$ of the binary correlator ${\cal W}(x)$.
The function ${\cal W}(R_c\rho)$ is the dimensionless correlator
of the dimensionless variable $\rho$, with the scale of decrease
of the order of $1$. As a result, the function ${\cal W}(R_c\rho)$
does not depend on $R_c$. Therefore, the integral over $\rho$
entering Eq.~(\ref{W(0)}) is a positive constant of the order of
unity. For example, in the case of the Gaussian correlations (see
Eq.~(\ref{WFT-Gaus})), we have, $W(0)=\sqrt{2\pi}R_c$, therefore,
the integral is $\sqrt{\pi/2}$.

For the second length in Eq.~\eqref{UB-Ln-sum} one gets,
\begin{subequations}\label{UB-Ln2-SSR}
\begin{eqnarray}
&&\frac{\Lambda_n}{2L^{(2)}_n}\approx
\pi^2\frac{(k\sigma)^4}{k_nd} \frac{n^3}{(kd/\pi)}\frac{T(0)}{k^3}
\nonumber\\[6pt]
&&\times\Bigg\{\left[\frac{1}{3}+\frac{1}{(2\pi n)^2}\right]^2
+\frac{16k_nd}{\pi^4(\pi n/d)^2}
\nonumber\\[6pt]
&&\times\left(\sum_{n'=1}^{n-1}+\sum_{n'=n+1}^{N_d}\right)
\frac{(\pi n'/d)^2}{k_{n'}d}\frac{(n^2+n'^2)^2}{(n^2-n'^2)^4}\Bigg\}
\qquad\label{UB-Ln2-SSR-Sum}\\[6pt]
&&\approx\frac{\pi^2}{4}\frac{(k\sigma)^4}{k_nd}
\frac{n^3}{(kd/\pi)}\frac{T(0)}{k^3}. \label{UB-Ln2-SSR-NT}
\end{eqnarray}
\end{subequations}
In Eq.~(\ref{UB-Ln2-SSR-Sum}) every term in the sum rapidly
decreases with an increase of the absolute value of $\Delta
n=n-n'$. This can be seen by making use of the following estimate,
\begin{equation}\label{factorDeltan}
\frac{(n^2+{n'}^2)^2}{(n^2-{n'}^2)^4}\approx\frac{1}{4(\Delta
n)^4}\qquad\mbox{for}\quad n\gg|\Delta n|.
\end{equation}
Therefore, the sum in Eq.~(\ref{UB-Ln2-SSR-Sum}) can be correctly
estimated by three terms with $n'=n,n\pm1$. For simplicity, in
Eq.~(\ref{UB-Ln2-SSR-NT}) we assume $N_d\gg n\gg 1$, and replace
the curly braces by factor $1/4$.

The explicit form for $T(0)$ directly follows from the definition
(\ref{T-def}) for the correlator $T(k_x)$,
\begin{eqnarray}\label{T(0)}
T(0)&=&\int_{-\infty}^{\infty}dx{{\cal W}''}^2(x)
\nonumber\\[6pt]
&=&R_c^{-3}\int_{-\infty}^\infty d\rho \left[\frac{d^2{\cal
W}(R_c\rho)}{d\rho^2}\right]^2.
\end{eqnarray}
If the roughness correlations are of the Gaussian form, then
according to Eq.~(\ref{T-Gaus-expl}), we have
$T(0)=3\sqrt{\pi}/4R_c^3$ and the integral over $\rho$ entering
Eq.~(\ref{T(0)}) is equal to $3\sqrt{\pi}/4$.

The second term in Eq.~\eqref{UB-Ln-sum} is bigger than the third
one. Therefore, a good aproximation for Eq.~\eqref{UB-Ln-sum} is
\begin{equation}\label{UB-Ln-sum-approx}
\frac{1}{L_n}\approx\frac{1}{L^{(1)}_n}+\frac{1}{L^{(2)}_n}.
\end{equation}
One can make sure of this if we consider in Eq.~\eqref{UB-Lnn'2lu}
the product of the fast decreasing factor \eqref{factorDeltan} and
$\cos^4[\pi(n-n')/2]$ (the latter cancels the terms with $\Delta
n=1,3,..$). Thus, we arrive at the following estimate,
\begin{equation}\label{L2updownarrow-approx}
\frac{\Lambda_n}{2L^{(2)\updownarrow}_n}
\approx \frac{\Lambda_n}{2L^{(2)\updownarrow}_{n,n}}
\approx\frac{\pi^2}{36}\frac{(k\sigma)^4}{k_nd}
\frac{n^3}{(kd/\pi)}\frac{T(0)}{k^3}.
\end{equation}
By comparing Eq.~\eqref{L2updownarrow-approx} with
Eq.~\eqref{UB-Ln2-SSR-NT} one can confirm
Eq.~\eqref{UB-Ln-sum-approx}.

\begin{figure}[t]
\includegraphics[angle=270,width=\columnwidth]
{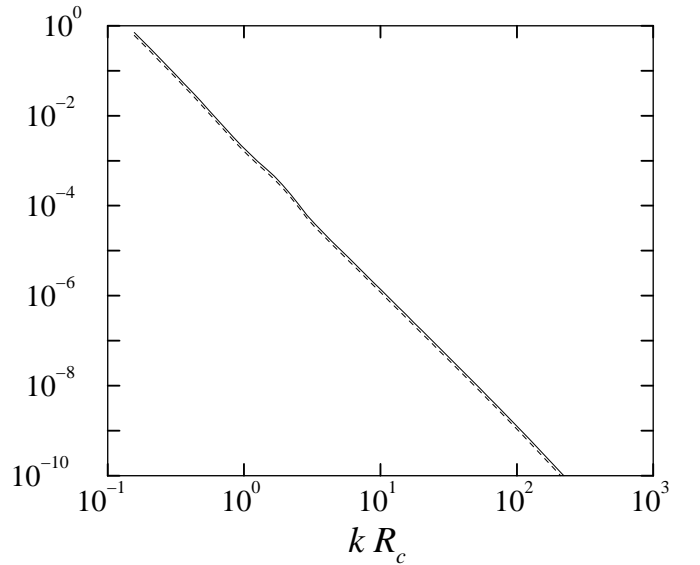} \caption{\label{fig:L2_VS_L2lu} Plot of
$\Lambda_n/2L_n^{(2)}$ (solid line) and $\Lambda_n/2L_n^{(2)}+
\Lambda_n/2L_n^{(2)\updownarrow}$ (dashed line) versus $kR_c$, for
$k d/\pi=100.5$, $n=50$ and $(k\sigma)^2=10^{-2}$.}
\end{figure}

The expression \eqref{UB-Ln-sum-approx} works well in the whole
domain of applicability of our theory, not only in the region of
small-scale roughness. To support this, in
Fig.~\ref{fig:L2_VS_L2lu} we compare the plot of
$\Lambda_n/2L_n^{(2)}$ with the plot of $\Lambda_n/2L_n^{(2)}+
\Lambda_n/2L_n^{(2)\updownarrow}$; they are practically the same.
Therefore, in what follows for this case of uncorrelated
boundaries we will neglect the third term in
Eq.~\eqref{UB-Ln-sum}.

Due to Eq.~\eqref{UB-Ln-sum-approx}, the relationship between the
first and second attenuation lengths is expressed by
\begin{equation}\label{UB-L1/L2-SSR}
\frac{L^{(1)}_n}{L^{(2)}_n}\sim\,(k\sigma)^2
\frac{n^2}{k_nd}\,(kR_c)^{-4}.
\end{equation}
According to a substantially different behavior of the quantities
$\Lambda_n/2L^{(1)}_n$ and $\Lambda_n/2L^{(2)}_n$ with respect to
$kR_c$, it becomes clear that they have to intersect at the
crossing point $(kR_c)_{cr}$. If the crossing point falls onto the
region of a small-scale roughness ($kR_c\ll1$), its dependence on
the model parameters is obtained by equating the expression
(\ref{UB-L1/L2-SSR}) to unity,
\begin{equation}\label{UB-cp-SSR}
(kR_c)_{cr}^2\sim (k\sigma)n/\sqrt{k_nd}.
\end{equation}
To the left from this point $(kR_c)_{cr}$ the SGS length prevails,
$L^{(2)}_n\ll L^{(1)}_n$. To its right the main contribution is
due to the first attenuation length, $L^{(1)}_n\ll L^{(2)}_n$. The
expression (\ref{UB-cp-SSR}) shows that the crossing point is
smaller for smaller values of $k\sigma$, as well as for smaller
mode indices $n$, or for larger values of the mode parameter
$kd/\pi$.

\subsubsection{Large-scale roughness: weak correlations}
\label{subsubsec-UB-Ln-WC}

The intermediate situation arises when the correlation length
$R_c$ becomes much larger than the wave length $2\pi/k$, but still
remains much less than the cycle length $\Lambda_n$,
\begin{equation}\label{UB-LSR-WC-def}
1\ll kR_c\ll k\Lambda_n.
\end{equation}
As before, the first of the weak-scattering conditions
(\ref{WS-Ln}) is the most restrictive, $R_c\ll\Lambda_n\ll2L_n$.

Since the distance $\Lambda_n$ is larger than the correlation
length $R_c$, the successive reflections of the waves from the
rough surface are weakly correlated. Note that the distance
between neighboring wave numbers $k_n$ and $k_{n\pm1}$ is much
smaller than the variation scale $R_c^{-1}$ of the correlators
$W(k_x)$ and $T(k_x)$,
\begin{equation}\label{kn-WC}
|k_n-k_{n\pm1}|\approx |\partial k_n/\partial n|=
\pi\Lambda_n^{-1}\ll R_c^{-1}.
\end{equation}
This implies that the correlators $W(k_x)$ and $T(k_x)$ are smooth
functions of the summation index $n'$. Therefore, the sum in the
expression (\ref{UB-Ln1-def}) for the first attenuation length
$L^{(1)}_n$ can be substituted by the integral,
\begin{subequations}\label{UB-Ln1-WC}
\begin{eqnarray}
\label{UB-Ln1-WC-W} &&\frac{\Lambda_n}{2L^{(1)}_n}\approx
(k\sigma)^2\frac{n}{kd/\pi}\, \int_{0}^{N_d} dn'\frac{(\pi
n'/d)^2}{k_{n'}d}
\nonumber\\[6pt]
&&\times\frac{1}{k}\left[W(k_n+k_{n'})+W(k_n-k_{n'})\right] \\[6pt]
\label{UB-Ln1-WC-fin}
&&=\frac{(k\sigma)^2}{\pi}\frac{n}{kd/\pi}\,\int_{-k}^{k}
dk_x\sqrt{k^2-k_x^2}\,\frac{W(k_n-k_x)}{k}.\qquad\quad
\end{eqnarray}
\end{subequations}
Eq.~(\ref{UB-Ln1-WC}) shows that the first attenuation length is
contributed by the scattering of a given $n$-th mode into other
propagating modes. In order to obtain this asymptotic result, we
have used only the condition of week correlations,
$R_c\ll\Lambda_n$. Therefore, Eq.~(\ref{UB-Ln1-WC}) provides a
reduction to Eq.~(\ref{UB-Ln1-SSR}) for small-scale corrugations,
$kR_c\ll 1$. In the case of a large-scale roughness, $kR_c\gg 1$,
the formula (\ref{UB-Ln1-WC}) obtained for $L^{(1)}_n$ allows one
for further simplifications as was done in
Ref.~\onlinecite{BFb79}.

In contrast with $L^{(1)}_n$, due to a rapidly decaying factor
(\ref{factorDeltan}) the SGS length $L^{(2)}_n$ can be still
described by keeping tree terms only, $n'=n,n\pm 1$, in the sum in
Eq.~(\ref{UB-Ln2}). Taking into account the estimate
(\ref{kn-WC}), for the case $N_d\gg n\gg 1$ one can obtain,
\begin{subequations}\label{UB-Ln2-WC}
\begin{eqnarray}
&&\frac{\Lambda_n}{2L^{(2)}_n}\approx\frac{\pi^2}{2}
\frac{(k\sigma)^4}{k_nd}\frac{n^3}{(kd/\pi)}\frac{T(0)+T(2k_n)}{k^3}
\nonumber\\[6pt]
&&\times\Bigg\{\left[\frac{1}{3}+\frac{1}{(2\pi n)^2}\right]^2
+\frac{8}{\pi^4}\Bigg\}
\label{UB-Ln2-WC-NT}\\[6pt]
&&\approx\frac{\pi^2}{8}\frac{(k\sigma)^4}{k_nd}
\frac{n^3}{(kd/\pi)}\frac{T(0)+T(2k_n)}{k^3}.
\qquad\quad\label{UB-Ln2-WC-fin}
\end{eqnarray}
\end{subequations}
To get the final expression (\ref{UB-Ln2-WC-fin}), we replaced the
curly braces in Eq.~(\ref{UB-Ln2-WC-NT}) by the factor $1/4$.
Naturally, at small-scale corrugations, $kR_c\ll 1$, the obtained
result (\ref{UB-Ln2-WC}) reduces to Eq.~(\ref{UB-Ln2-SSR}). For a
large-scale roughness, $kR_c\gg 1$, one should use
Eq.~(\ref{UB-Ln2-WC}) because of an arbitrary value of the
parameter $k_nR_c$. We do not consider here this case in detail
due to its intermediate character.

\subsubsection{Large-scale roughness: strong correlations}
\label{subsubsec-UB-Ln-SC}

Another limit case refers to the correlation length $R_c$ to be
very large both in comparison with the wave length $2\pi/k$ and
with the cycle length $\Lambda_n$,
\begin{equation}\label{LSR-SC-def}
1\lesssim k\Lambda_n\ll kR_c.
\end{equation}
In this case the number of wave reflections on the scale of the
correlation length $R_c$ is large. Therefore, the successive
reflections are strongly correlated to each other.

Under the relations (\ref{LSR-SC-def}) the second of the
weak-scattering conditions (\ref{WS-Ln}) is the most restrictive,
therefore, the condition of the applicability reads as
\begin{equation}\label{domain-SC}
\Lambda_n\ll R_c\ll2L_n.
\end{equation}
The latter requirement determines the upper limit for the value of
the correlation length $R_c$.

Due to Eq.~(\ref{LSR-SC-def}), the distance between neighboring wave
numbers $k_n$ and $k_{n\pm1}$ turns out to be much larger than the
variation scale $R_c^{-1}$ of the correlators $W(k_x)$ and $T(k_x)$,
\begin{equation}\label{kn-SC}
|k_n-k_{n\pm1}|\approx|\partial k_n/\partial n|=\pi\Lambda_n^{-1}\gg
R_c^{-1}.
\end{equation}
This indicates that the probability of the intermode ($n'\neq n$)
transitions is exponentially small and the attenuation lengths are
mainly formed by the incoherent intramode ($n'=n$) scattering.
Formally, for strong correlations (\ref{LSR-SC-def}) the
correlators $W(k_x)$ and $T(k_x)$ are sharpest functions of the
summation index $n'$. In the sums of Eqs.~(\ref{UB-Ln1-def}) and
(\ref{UB-Ln2}) for the attenuation lengths the main contribution
is due to the diagonal terms with $n'=n$, for which $W(2k_n)$ and
$T(2k_n)$ can be neglected in comparison with $W(0)$ and $T(0)$.

Thus, the first attenuation length reads
\begin{equation}\label{UB-Ln1-SC}
\frac{\Lambda_n}{2L^{(1)}_n}\approx
\frac{(k\sigma)^2}{k_nd}\frac{n^3}{(kd/\pi)^3}\,kW(0).
\end{equation}
Correspondingly, for the SGS length one gets,
\begin{equation}\label{UB-Ln2-SC}
\frac{\Lambda_n}{2L^{(2)}_n}\approx\frac{\pi^2}{2}
\frac{(k\sigma)^4}{k_nd}\frac{n^3}{(kd/\pi)}\frac{T(0)}{k^3}
\left[\frac{1}{3}+\frac{1}{(2\pi n)^2}\right]^2.
\end{equation}

The ratio of the first attenuation length to the second one can be
presented as
\begin{equation}\label{L1/L2-SC<<1}
\frac{L^{(1)}_n}{L^{(2)}_n}\sim\,\frac{\Lambda_n}{2L_n^{(1)}}\,
\left(\frac{\Lambda_n}{R_c}\right)^5(k_n \Lambda_n)^{-2}\ll1.
\end{equation}
According to the first inequality in Eq.~(\ref{WS-Ln}), applied to
the condition (\ref{LSR-SC-def}) of the strong-correlations and to
the evident relationship $k_n\Lambda_n\gtrsim1$, one can see that
the amplitude scattering length always prevails over the SGS
length within the interval of strong correlations. For this reason
in the condition of applicability (\ref{domain-SC}) one should
substitute $L_n$ by $L^{(1)}_n$. This allows one to arrive at the
inequalities in the explicit form,
\begin{equation}\label{VR-SC}
k\Lambda_n \ll kR_c \ll (k\Lambda_n) (k\sigma)^{-1} (kd/\pi n).
\end{equation}

Before we pass to the next configuration of the waveguide, it is
interesting to compare the case of two uncorrelated boundaries
with the waveguide with one rough boundary, studied in
Ref.~\cite{IzMkRn-PRB-06}. One can found that under the
approximation \eqref{UB-Ln-sum-approx}, the expression for $1/L_n$
turns out to be as twice as bigger than that one for the
one-rough-boundary configuration. This fact is quite natural since
in the case of uncorrelated boundaries the two rough surfaces are
of the same statistical nature (see the factor $2$ in the first
two terms of Eq.~\eqref{UB-Q}).

The remarkable similarity between the behavior of the inverse
attenuation length for both configurations can be observed in
Fig.~\ref{fig:Q1D-2RB_L1L2_n_kRc}. Note that the normalization
factor $\Lambda_n$ for one boundary is twice as much of the
quantity for two boundaries, and both curves are practically
superposed.

\subsection{Waveguide with antisymmetric rough boundaries}
\label{subsec-CCB-L}
%
\begin{figure}[t]
\includegraphics[angle=270,width=\columnwidth]
{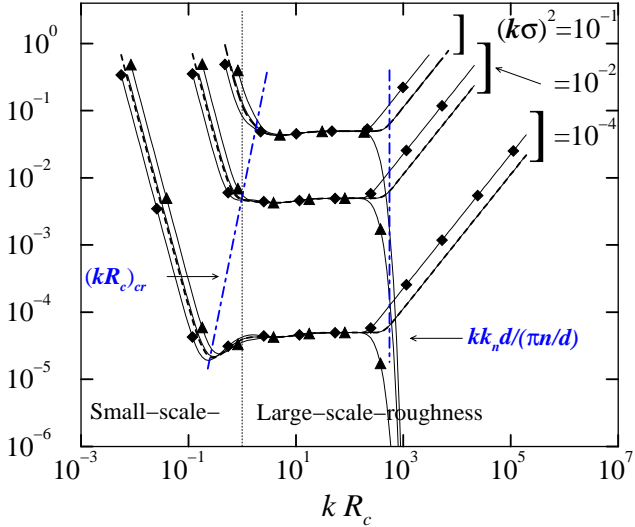} \caption{\label{fig:Q1D-2RB_L1L2_n_kRc} Plot of
$\Lambda_n/2L_n$ versus $kR_c$ for $k d/\pi=100.5$, $n=50$ and
different values of $(k\sigma)^2$. Different sets of curves are
shown for three waveguide configurations, as well as for the
waveguide with one rough boundary analyzed in
Ref.~\cite{IzMkRn-PRB-06} (dashed line). Curves without symbols
depict Eq.~\eqref{UB-Ln-sum} for the waveguide with uncorrelated
boundaries; curves labelled by triangles and diamonds stand for
Eqs.~\eqref{CCB-Ln-sum} and \eqref{SB-Ln-sum} for waveguides with
antisymmetric and symmetric boundaries, respectively}.
\end{figure}

According to Eqs.~(\ref{CCB-Q}) and (\ref{Ln-Q}), the inverse
attenuation length can be presented as a sum of the partial
inverse lengths $1/L_{n,n'}$ that are responsible for the
scattering from $n$th into $n'$th modes,
\begin{equation}\label{CCB-Ln}
\frac{1}{L_n}=\sum_{n'=1}^{N_d}\frac{1}{L_{n,n'}}.
\end{equation}
However, it is noteworthy that in this case the partial terms
describing the intramode $1/L_{n,n}$ and intermode $1/L_{n,n'\neq
n}$ scattering are rather different since they are associated with
different scattering mechanisms. Indeed, the intermode scattering
results from the RHP spectrum only, through the expression in
Eq.~(\ref{CCB-Q1}) for the kernel $Q_1(k_x,q_x)$. These partial
lengths are formed by the gradient scattering mechanism, due to
the term containing in the scattering potential the derivative
$\xi'(x)$, see Eq.~\eqref{CCB-Xi1-def}. They read as
\begin{eqnarray}\label{CCB-Lnn'}
\frac{1}{L_{n,n'\neq n}}&=&4\sigma^2\frac{(\pi n/d)^2}{k_nd}
\frac{(\pi n'/d)^2}{k_{n'}d}\,\sin^4[\pi(n-n')/2]
\nonumber\cr\\
&&\times[W(k_n+k_{n'})+W(k_n-k_{n'})].
\end{eqnarray}

From the above expression one can draw an interesting conclusion.
As one can see, in the case of symmetric boundaries the
transitions are forbidden between the modes for which the
difference of the indexes $n-n'$ is an even number. There are
transitions only between odd and even channels. This fact can be
explained by an existence of an additional integral of motion,
namely, the parity. Specifically, if one solves the stationary
problem for the effective Hamiltonian, there are two sets of
eigenstates of different parity. Correspondingly, there are two
independent sets of eigenvalues, see
Refs.~\cite{LunMenIz01,MnLnIz2004}.

The intramode scattering, related to the RSGP spectrum through the
kernel $Q_2(k_x,q_x)$ in the Eq.~\eqref{CCB-Q2}, is related to the
SGS mechanism via the function $\hat{\cal V}(x)$ in
Eq.~\eqref{CCB-Xi2-def}. Therefore, we have
\begin{equation}\label{CCB-Lnn}
\frac{1}{L_{n,n}}=\frac{\sigma^4}{2}\frac{(\pi n/d)^4}{k_n^2}
\left[T(2k_n)+T(0)\right].
\end{equation}
Note that in a single-mode waveguide with $N_d=1$ the sum over
$n'$ in Eq.~(\ref{CCB-Ln}) contains only one term with $n'=n=1$.
In this case the backscattering length $L_{1,1}^{(b)}$ entering in
Eq.~(\ref{CCB-Lnn}) is in accordance with that obtained in
Ref.~\cite{MT98}.

Bearing in mind very different dependencies of the $W(k_x)$ and
$T(k_x)$ power spectra on the roughness correlation length $R_c$,
one can write the inverse attenuation length as the sum of two
terms. One is related to $W(k_x)$,
\begin{equation}\label{CCB-Ln1}
\frac{1}{L^{(1)}_n}=\sum_{{\scriptstyle \begin{array}{c} n'=1,\\
n'\neq n \end{array}}}^{N_d}\frac{1}{L_{n,n'}},
\end{equation}
and the other is just the inverse of the intramode attenuation
length determined by Eq.~\eqref{CCB-Lnn}. Therefore,
$1/L_n^{(2)}=1/L_{n,n}$, and one can write,
\begin{equation}\label{CCB-Ln-sum}
\frac{1}{L_n}=\frac{1}{L^{(1)}_n}+\frac{1}{L^{(2)}_{n}}.
\end{equation}

In Fig.~\ref{fig:Q1D-2RB_L1L2_n_kRc} we depict the length defined
by Eq.~\eqref{CCB-Ln-sum} as a function of $kR_c$ for different
values of $k\sigma$; the curve is labelled with triangles.

As above, it is convenient to consider separately three regions of
the correlation length.

\subsubsection{Small-scale roughness}
\label{subsubsec-CCB-Ln-SSR}

We start with the standard case of a small-scale boundary
perturbation specified by the conditions \eqref{SSR-def}. In this
case for any term in the summation over $n'$ the argument of the
correlators $W(k_x)$ and $T(k_x)$ is much less than the scale of
their decrease $R_c^{-1}$, therefore, it can be taken as zero.
Thus, for the attenuation length describing the intermode
scattering one gets,
\begin{subequations}\label{CCB-Ln1-SSR}
\begin{eqnarray}
\frac{\Lambda_n}{2L^{(1)}_n}&\approx&
4(k\sigma)^2\frac{n}{kd/\pi}\frac{W(0)}{k}\nonumber \\[6pt]
&&\times\sum_{n'=1}^{N_d}\frac{(\pi n'/d)^2}{k_{n'}d}
\sin^4[\pi(n-n')/2]\qquad
\label{CCB-Ln1-SSR-Sum1}\\[6pt]
&\approx& (k\sigma)^2\frac{n}{kd/\pi}\,\frac{kW(0)}{2}
\label{CCB-Ln1-SSR-W}
\end{eqnarray}
\end{subequations}
The factor $\sin^4[\pi(n-n')/2]$ gives alternating values of one
and zero for $n-n'=1,3,..$ and $n-n'=2,4,..$, respectively,
therefore, for a large number of conducting modes $N_d\gg 1$ it
can be replaced by $1/2$. The estimate (\ref{CCB-Ln1-SSR-W}) is
obtained with the change of the summation by integration. For a
further analysis one has to take into account the formula in
Eq.~\eqref{W(0)} for $W(0)$.

The second term in Eq.~\eqref{CCB-Ln-sum} reads as
\begin{equation}\label{CCB-Ln2-SSR}
\frac{\Lambda_n}{2L^{(2)}_n}\approx \frac{\pi^2}{2}
\frac{(k\sigma)^4}{k_nd}\frac{n^3}{(kd/\pi)}\frac{T(0)}{k^3}\,,
\end{equation}
where the explicit form for $T(0)$ can be taken as in
Eq.~\eqref{T(0)}.

It is easy to realize that the relationship between lengths
defined by Eqs.~\eqref{CCB-Ln1-SSR-W} and \eqref{CCB-Ln2-SSR} is
of the same order as in the case of uncorrelated boundaries, see
Eq.~\eqref{UB-L1/L2-SSR}. In this way one obtains the expression
\eqref{UB-cp-SSR} for the crossing point $(kR_c)_{cr}$ at which
the curves of $\Lambda_n/2L^{(1)}_n$ and $\Lambda_n/2L^{(2)}_n$
intercept each other.

\subsubsection{Large-scale roughness: weak correlations}
\label{subsubsec-CCB-Ln-WC}

This intermediate situation is specified by Eq.~
\eqref{UB-LSR-WC-def}. In Eq.~\eqref{CCB-Ln1} we replace
$\sin^4[\pi(n-n')/2]$ by $1/2$, and change the sum by the integral
since the correlator $W(k_x)$ is a smooth function of $n'$ (see
the discussion before Eq.~\eqref{kn-WC}),
\begin{equation}\label{CCB-Ln1-WC}
\frac{\Lambda_n}{2L^{(1)}_n}\approx
\frac{(k\sigma)^2}{\pi}\frac{n}{kd/\pi}\,\int_{-k}^{k}dk_x
\sqrt{k^2-k_x^2}\,\frac{W(k_n-k_x)}{k}.
\end{equation}
Note that Eq.~\eqref{CCB-Ln1-WC} coincides with
Eq.~\eqref{UB-Ln1-WC-fin} for the case of two uncorrelated
boundaries.

For the length $L^{(2)}_n$ in the large-scale-roughness region,
$kR_c\gg 1$, one should use the general expression (\ref{CCB-Lnn})
because of arbitrary value of the parameter $k_nR_c$. Its
normalized version reads
\begin{equation}\label{CCB-Lnn-WC}
\frac{\Lambda_n}{2L_{n,n}}=\frac{\pi^2}{4}\frac{(k\sigma)^4}{k_nd}
\frac{n^3}{kd/\pi}\frac{T(2k_n)+T(0)}{k^3}.
\end{equation}

\subsubsection{Large-scale roughness: strong correlations}
\label{subsubsec-CCB-Ln-SC}

In this limit case specified by Eq.~\eqref{LSR-SC-def}, the
distance between wave numbers $k_n$ and $k_{n\pm1}$ is much larger
than the variation scale $R_c^{-1}$ of the correlator $W(k_x)$,
see Eq.~\eqref{kn-SC}. This means that the probability of the
intermode ($n'\neq n$) transitions is exponentially small and the
total attenuation length is mainly formed by the incoherent
intramode ($n'=n$) scattering,
\begin{equation}\label{CCB-Ln-approx}
\frac{1}{L_n}\approx \frac{1}{L^{(2)}_{n}}.
\end{equation}
For strong correlations the correlator $T(k_x)$ is a sharp
function of its argument $k_x$. Thus, the term with $T(2k_n)$ can
be neglected in comparison with $T(0)$, and one gets,
\begin{equation}\label{CCB-Ln2-reg3-SSR}
\frac{\Lambda_n}{2L_n}\approx \frac{\pi^2}{4}
\frac{(k\sigma)^4}{k_nd} \frac{n^3}{(kd/\pi)}\frac{T(0)}{k^3}.
\end{equation}

In this region of $kR_c$ the SGS mechanism is recovered,
determining the total attenuation length. The second inequality in
Eq.~\eqref{domain-SC} is automatically satisfied for any value of
$kR_c$

\subsection{Waveguide with Symmetric Rough Boundaries}
\label{subsec-SB-L}

As above, for this configuration it is convenient to write down
the inverse attenuation lengths as a sum of two terms,
\begin{equation}\label{SB-Ln-sum}
\frac{1}{L_n}=\frac{1}{L^{(1)}_n}+\frac{1}{L^{(2)}_{n}}.
\end{equation}
Here each term is associated with the corresponding term in
Eq.~\eqref{SB-Q}, resulting from different mechanisms of the
surface scattering.

The first term in Eq.~\eqref{SB-Ln-sum} is related to the RHP
spectrum $W(k_x)$ through the expression for $Q_1(k_x,k_x')$.
According to Eq.~(\ref{SB-Q1}), we have
\begin{eqnarray}\label{SB-Ln1-def}
\frac{1}{L^{(1)}_n}&=&
4\sigma^2\,\frac{(\pi n/d)^2}{k_nd}\,\sum_{n'=1}^{N_d}
\frac{(\pi n'/d)^2}{k_{n'}d}\cos^4[\pi(n-n')/2]\nonumber\\[6pt]
&&\times\left[ W(k_n+k_{n'})+W(k_n-k_{n'})\right].\qquad\qquad
\end{eqnarray}
The term $1/L^{(1)}_{n,n}$ corresponding to $n'=n$,  is related to
the amplitude scattering while the terms $1/L^{(1)}_{n,n'\neq n}$
result from the gradient scattering. These two mechanisms of
surface scattering are due to the corresponding terms in the
expression for $\Xi_{1}(k_x,k_x')$, see Eq.~\eqref{SB-Xi1}.
Namely, the former follows from the term depending on the
amplitude of the roughness profile $\sigma\xi_i(x)$, and the
latter from the terms depending on the roughness gradient
$\sigma\xi_i'(x)$.

The second term in Eq.~\eqref{SB-Ln-sum} related to the RSGP
spectrum through $Q_2(k_x,k_x')$, is associated solely with the
SGS mechanism due to the square-gradient function $\widehat{\cal
V}(x)$, see Eq.~(\ref{SB-Xi2}). In accordance with
Eq.~(\ref{SB-Q2}), it is described by
\begin{equation}\label{SB-Ln2}
\frac{1}{L^{(2)}_n}=\sum_{n'=1}^{N_d}\frac{1}{L^{(2)}_{n,n'}}.
\end{equation}
Here
\begin{equation}\label{SB-Lnn2}
\frac{1}{L^{(2)}_{n,n}}=\frac{\sigma^4}{2}\,\frac{(\pi n/d)^4}{k_n^2}\,
\left[\frac{1}{3}+\frac{1}{(\pi n)^2}\right]^2 \left[T(2k_n)+T(0)\right]
\end{equation}
and
\begin{equation}\label{SB-Lnn'2}
\begin{split}
\frac{1}{L^{(2)}_{n,n'\neq n}}=&\frac{128\sigma^4}{\pi^4}\frac{(\pi
n/d)^2}{k_n}\,\frac{(\pi n'/d)^2}{k_{n'}}\,
\frac{(n^2+n'^2)^2}{(n^2-n'^2)^4}\\[6pt]
&\times\cos^4[\pi(n-n')/2]\\[6pt]
&\times\left[T(k_n+k_{n'})+T(k_n-k_{n'})\right].
\end{split}
\end{equation}
Note that in a single-mode waveguide with $N_d=1$ (therefore with
$n'=n=1$ in Eqs.~\eqref{SB-Ln1-def} and \eqref{SB-Ln2}) the
backscattering length $L_{11}^{(b)}$ takes the expression derived
in Ref.~\cite{MT01}.

The diagonal part of the inverse attenuation length,
$1/L_{n,n}=1/L^{(1)}_{n,n}+1/L^{(2)}_{n,n}$, characterizes the
intramode scattering. The off-diagonal partial attenuation length
$L_{n\neq n'}=1/L^{(1)}_{n,n'\neq n}+1/L^{(2)}_{n,n'\neq n}$
describes the intermode scattering.

To compare with the above case of the antisymmetric boundaries,
here there are no transitions between the modes with odd values
for the difference $n-n'$. Therefore, only the transitions between
odd or even modes are allowed only. This  fact is related to the
underlying symmetry, and can be associated with specific integrals
of motion of the total Hamiltonian.

\subsubsection{Small-scale roughness}
\label{subsubsec-SB-Ln-SSR}

In the case of a small-scale roughness the first attenuation
length reads
\begin{subequations}\label{SB-Ln1-SSR}
\begin{eqnarray}
&&\frac{\Lambda_n}{2L^{(1)}_n}\approx
2(k\sigma)^2\frac{n}{kd/\pi}\frac{W(0)}{k}\sum_{n'=1}^{N_d}
\frac{(\pi n'/d)^2}{k_{n'}d}\qquad
\label{SB-Ln1-SSR-Sum}\\[6pt]
&&\approx (k\sigma)^2\frac{n}{kd/\pi}\,\frac{kW(0)}{2}.
\label{SB-Ln1-SSR-W}
\end{eqnarray}
\end{subequations}

For the second length we have,
\begin{subequations}\label{SB-Ln2-SSR}
\begin{eqnarray}
&&\frac{\Lambda_n}{2L^{(2)}_n}\approx
\frac{\pi^2}{2}\frac{(k\sigma)^4}{k_nd} \frac{n^3}{(kd/\pi)}
\frac{T(0)}{k^3} \nonumber\\[6pt]
&&\times\Bigg\{\left[\frac{1}{3}+\frac{1}{(\pi n)^2}\right]^2
+\frac{256k_nd}{\pi^4(\pi n/d)^2}
\nonumber\\[6pt]
&&\times\left(\sum_{n'=1}^{n-1}+\sum_{n'=n+1}^{N_d}\right)
\frac{(\pi n'/d)^2}{k_{n'}d}\frac{(n^2+n'^2)^2}{(n^2-n'^2)^4}
\nonumber\\[6pt]
&&\times\cos^4[\pi(n-n')/2]\Bigg\}
\qquad\label{SB-Ln2-SSR-Sum}\\[6pt]
&&\approx\frac{\pi^2}{8}\frac{(k\sigma)^4}{k_nd}
\frac{n^3}{(kd/\pi)}\frac{T(0)}{k^3}.\label{SB-Ln2-SSR-fin}
\end{eqnarray}
\end{subequations}
As well as in the configuration with uncorrelated boundaries, due
to the estimate \eqref{factorDeltan} every term in the sum of
Eq.~(\ref{SB-Ln2-SSR-Sum}) rapidly decreases with an increase of
the absolute value of $\Delta n=n-n'$. In addition, the factor
$\cos^4\pi(n-n')/2$ is zero for $n-n'=1,3,..$. Therefore, the sum
in Eq.~(\ref{SB-Ln2-SSR-Sum}) can be evaluated by three terms with
$n'=n,n\pm2$, and one can replace the quantity in the curly braces
by $1/4$. Here we also assume, $N_d\gg n\gg 1$.

It is evident that in the small-roughness case the crossing point
is determined by Eq.~\eqref{UB-cp-SSR}.

\subsubsection{Large-scale roughness: weak correlations}
\label{subsubsec-SB-Ln-WC}

In the intermediate situation \eqref{UB-LSR-WC-def} the first
attenuation length $L^{(1)}_n$ coincides with that obtained for
the previous configurations,
\begin{equation}\label{SB-Ln1-WC-fin}
\frac{\Lambda_n}{2L^{(1)}_n}\approx
\frac{(k\sigma)^2}{\pi}\frac{n}{kd/\pi}\,\int_{-k}^{k}
dk_x\sqrt{k^2-k_x^2}\,\frac{W(k_n-k_x)}{k}.
\end{equation}
As for the second length $L^{(2)}_n$, for $N_d\gg n\gg 1$ it can
be written,
\begin{equation}\label{SB-Ln2-WC-fin}
\frac{\Lambda_n}{2L^{(2)}_n}\approx\frac{\pi^2}{16}
\frac{(k\sigma)^4}{k_nd}\frac{n^3}{(kd/\pi)}\frac{T(0)+T(2k_n)}{k^3}.
\end{equation}

\subsubsection{Large-scale roughness: strong correlations}
\label{subsubsec-SB-Ln-SC}

In this region of $kR_c$ the first attenuation length reads,
\begin{equation}\label{SB-Ln1-SC}
\frac{\Lambda_n}{2L^{(1)}_n}\approx
2\frac{(k\sigma)^2}{k_nd}\frac{n^3}{(kd/\pi)^3}\,kW(0).
\end{equation}
For the SGS length one gets,
\begin{equation}\label{SB-Ln2-SC}
\frac{\Lambda_n}{2L^{(2)}_n}\approx\frac{\pi^2}{4}
\frac{(k\sigma)^4}{k_nd}\frac{n^3}{(kd/\pi)}\frac{T(0)}{k^3}
\left[\frac{1}{3}+\frac{1}{(\pi n)^2}\right]^2.
\end{equation}

The ratio of the first attenuation length to the second one can be
obtained due to Eq.~\eqref{L1/L2-SC<<1}. One can see that the
amplitude scattering length always prevails over the SGS length
within the interval of strong correlations,
$L_n^{(1)}/L_n^{(2)}\ll 1$. For this reason, the condition of
applicability coincides with Eq.~\eqref{VR-SC}.

In Fig.~\ref{fig:Q1D-2RB_L1L2_n_kRc} by diamonds we show the
behavior of the total dimensionless inverse scattering length as a
function of the correlation parameter $kR_c$.

\section{Numerical analysis of the surface
scattering potential $\widehat{U}$} \label{sec-num-analysis}

In this Section we present numerical data for periodic waveguides
(billiards) with the potential $\widehat{U}(x,z)$. Our main
attention is payed to the structure of eigenstates for the
discussed types of boundaries. Although this information is not
directly related to the scattering properties of open waveguides,
it is instructive to reveal a role of the SGS-terms in the
potential, by comparing the structure of eigenstates for different
boundaries. The numerical method is described in details in
Ref.\cite{MnLnIz2004}; below we discuss the most interesting
results and relate them to the previous studies (see,
Ref.\cite{LunMenIz01} and references therein).

Our particular interest is in the ``repulsion'' effect discovered
in Ref.~\cite{MnLnIz2004} for quasi-1D billiards with rough
surfaces. This phenomena can be observed for a subset of
eigenfunctions with the smallest values of the transversal quantum
number of the unperturbed billiard with perfectly flat boundaries.
For these eigenstates it was found that the maximum of the
eigenfunction intensity is strongly shifted towards from the rough
boundary. This ``repulsion"  was shown to increase with a decrease
of the roughness correlation length $R_c$. Based on our numerical
data, one can make a conclusion that this effect of ``repulsion"
is directly related to the SGS-terms discussed above.

To start with, we show in Fig.~\ref{fig:Psi(x,z)-SGS-noInterSGS} a
typical eigenstate of this type for the billiard with one rough
surface (upper panel). To compare with, the low panel presents the
corresponding eigenstate for the same potential, however, without
SGS-terms. As one can see, by removing these terms, the repulsion
disappears.

It is also instructive to compare the structure of such
eigenstates with those computed for the case of both symmetric and
antisymmetric rough boundaries, see
Fig.~\ref{fig:Psi(x,z)-AntiSym-Sym}. In both cases the potentials
have the SGS-terms. However, the repulsion phenomena is absent in
the antisymmetric billiard. A close inspection reveals that the
difference between these two cases is entirely due to the fact
that in the antisymmetric billiard there is no square-gradient
scattering between different channels, in contrast to the
symmetric billiard. Thus one can make a conclusion that the main
contribution to the repulsion is due to the intermode SGS-terms.
This conclusion supports an observation found in
Ref.~\cite{MnLnIz2004} according to which the effect of the
repulsion can be explained due to a strong localization in the
channel space. This localization occurs due to a relatively strong
interaction between different conducting channels (waveguide
modes).

\begin{figure}[t]
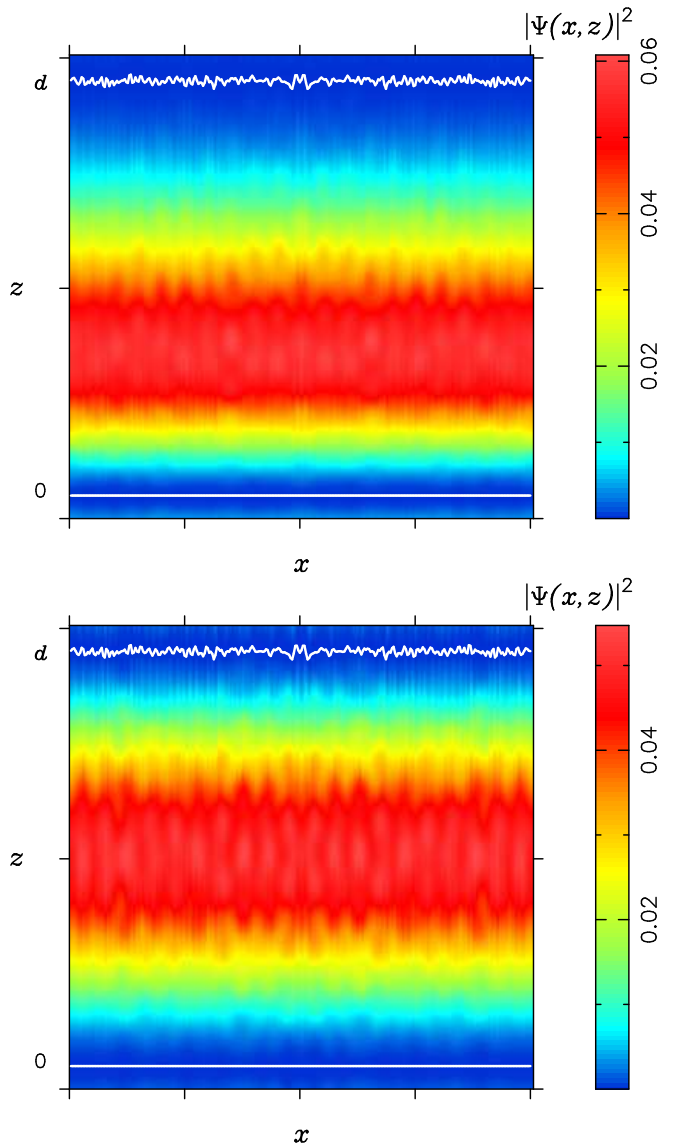

\includegraphics[angle=270,width=\columnwidth]
{fig3.ps}
\includegraphics[angle=270,width=\columnwidth]
{fig4.ps} \caption{\label{fig:Psi(x,z)-SGS-noInterSGS} (Color
online) Upper panel: intensity of a typical eigenstate manifesting
a ``repulsion" from the rough surface. Low panel: the
corresponding eigenstate for the potential without the SGS-terms.}
\end{figure}

\begin{figure}[t]
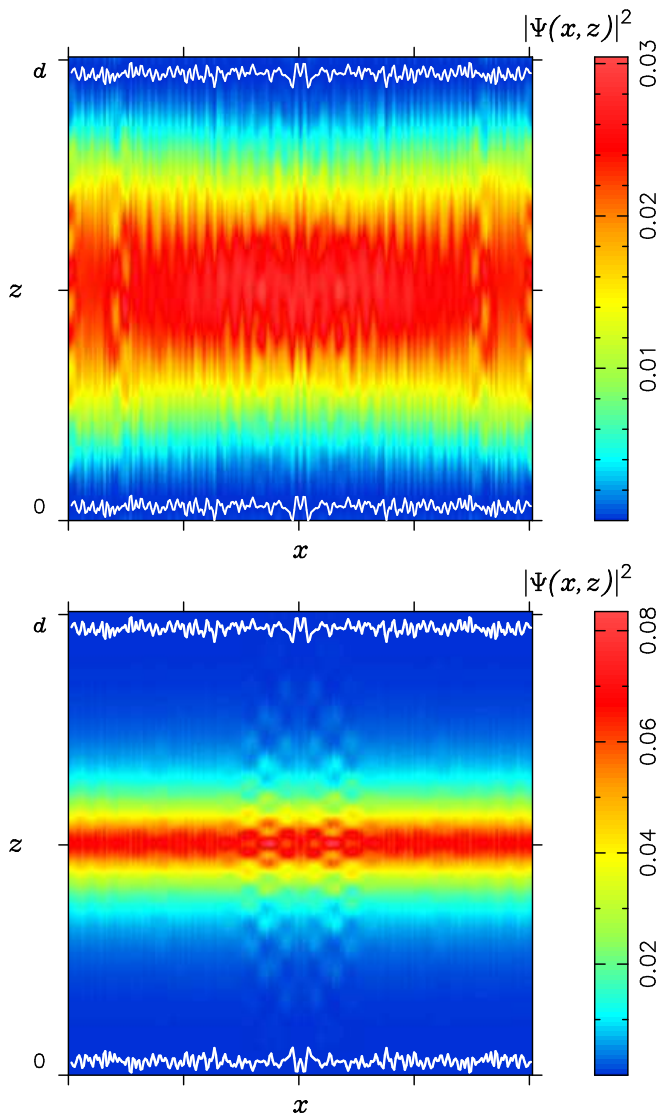

\includegraphics[angle=270,width=\columnwidth]
{fig5.ps}
\includegraphics[angle=270,width=\columnwidth]
{fig6.ps} \caption{\label{fig:Psi(x,z)-AntiSym-Sym} (Color online)
Intensity of an eigenfunction revealing strong repulsion from two
symmetric rough boundaries (low panel). The same for the
antisymmetric billiard for which there is no interaction between
different channels (upper panel).}
\end{figure}

\section{Conclusion}
\label{Sec-Conclusion}

We have developed a perturbative approach to the surface
scattering in multimode quasi-1D waveguides with rough surfaces.
The approach is based on the construction of an effective
Hamiltonian that formally corresponds to the ``bulk" scattering
described by the potential of a very complicated structure. The
detailed analysis of this potential has led to the conclusion that
there are specific square-gradient scattering (SGS) terms that
were not taken into account in the previous studies. We show that
in spite of their seeming smallness in comparison with the
standard terms, they can give a major contribution to the value of
the scattering length in the case of a sufficiently strong
roughness characterized by a corresponding value of the derivative
of boundary profiles.

In order to demonstrate a strong influence of this SGS-mechanism,
we have applied the approach to the waveguides with the boundaries
of three different types. Specifically, we considered the
waveguides with two uncorrelated boundaries, and compared the
results with those obtained for other two cases of symmetric and
antisymmetric rough boundaries. In this way one can see the
influence of various terms very clearly.

According to our analysis, all terms in the expression for the
effective potential can be regrouped in three parts. The first one
is due to the roughness mean amplitude $\sigma$. Another group is
characterized by the derivative of profiles, reflecting the fact
that the scattering depends on two characteristic parameters,
namely, on the amplitude $\sigma$ (or, the variance $\sigma^2$),
and on the correlation length $R_c$ inversely proportional to the
mean derivative of the profiles. In the expression for the
scattering potential the above two types of terms have linear
dependence on the root-mean-square roughness height $\sigma$.
There is one more group of terms that can not be reduced to the
two previous ones, and that gives rise to the square dependence on
the profile derivative $\sigma/R_c$. For this reason, in the
previous studies it was assumed that these SGS-terms can be safely
neglected.

However, we show that apart from the $\sigma$-dependence due to
the amplitude and gradient terms, as well as the
$\sigma^2$-dependence due to the SGS-terms, there is also
dependence on the correlation parameter $R_c$. As was found, the
$R_c$-dependence for the SGS-terms turns out to be very different
from those corresponding to the amplitude and gradient terms. Our
analysis reveals that there is a region of parameters where the
SGS-contribution to the inverse scattering length is compared (or
even larger) with the standard contributions due to the amplitude
and gradient scattering.

In accordance with the general theory of surface scattering
\cite{BFb79} the wave propagation through any $n$-th conducting
channel is determined by its total (or outgoing) attenuation
length $L_n$. The inverse attenuation length $1/L_n$ can be
presented as the sum of partial inverse lengths, see
Eqs.~\eqref{UB-Ln-sum}, \eqref{CCB-Ln-sum} and \eqref{SB-Ln-sum}.
The first term $1/L_n^{(1)}$ is associated with both the amplitude
and gradient scattering, it depends on the roughness-height power
spectrum $W(k_x)$. To the contrary, the partial length $L_n^{(2)}$
being connected with the SGS-terms, depends on the power spectrum
$T(k_x)$ that is principally different from $W(k_x)$. For the
convenience of comparison between three different types of
boundaries, in Eq.~\eqref{UB-Ln-sum} we have separately written
the term $1/L_n^{(2)\updownarrow}$ that appears in the specific
case of two uncorrelated boundaries. This term also depends on
$T(k_x)$.

Such a representation allows us to analyze the total dependence of
the inverse attenuation length $1/L_n$ on two control parameters,
the roughness hight $\sigma$ and the correlation length $R_c$. Our
analysis shows that one can speak about three characteristic regions
of $R_c$, namely, the small-scale \eqref{SSR-def}, large-scale
weak-correlation \eqref{UB-LSR-WC-def} and large-scale
strong-correlation \eqref{LSR-SC-def} roughness.

As we noticed above, the $W(k_x)$- and $T(k_x)$-power spectra have
very different dependencies on the roughness correlation length
$R_c$. Specifically, the inverse value of the $L_2$-length
decreases with an increase of the correlation parameter $kR_c$;
this occurs in the whole interval of the roughness values. As for
the inverse value of the $L_1$-length, it shows a different
behavior in each of the three regions of roughness and strongly
depends on the waveguide configuration.

In the region of a small-scale roughness \eqref{SSR-def} the
inverse length $1/L_1$ increases with an increase of $kR_c$.
Therefore, the curves corresponding to the terms $1/L_n^{(1)}$ and
$1/L_n^{(2)}$ ($1/L_n^{(2)}+1/L_n^{(2)\updownarrow}$ for the case
of uncorrelated boundaries) intersect each other, see Fig.~2. If
the crossing point $(kR_c)_{cr}$ falls into this region of
small-scale roughness, it obeys the law given by
Eq.~(\ref{UB-cp-SSR}). To the left from the crossing point
$(kR_c)_{cr}$, the SGS-length $L_2$ prevails over the standard
length, $L^{(2)}_n\ll L^{(1)}_n$. Contrary, to the right from the
point $(kR_c)_{cr}$, the $L_1$-length gives the main contribution
to the scattering process, $L^{(1)}_n\ll L^{(2)}_n$.

The detailed analysis of three types of boundaries shows that for
the small-scale roughness region one can specify that the
crossover (with an increase of $kR_c$) occurs due to the following
interplays:
\begin{itemize}
\item
uncorrelated boundaries: the two SGS-terms that depend on the
$T(k_x)$-power spectrum, versus the amplitude and gradient terms
depending on the $W(k_x)$-power spectrum;
\item
antisymmetric boundaries: the SGS-term versus the gradient term;
\item
symmetric boundaries: the SGS-term versus the amplitude and gradient
terms.
\end{itemize}

Thus, we show that in the region of small-scale roughness and for
any fixed value of the roughness height $\sigma$, one can indicate
the region of small values of the correlation length $R_c$ where
the SGS-length $L^{(2)}_{n}$ predominates over the the length
$L^{(1)}_{n}$. This predominance occurs in spite of the fact that
$1/L^{(1)}_{n}$ is proportional to $\sigma^2$ while
$1/L^{(2)}_{n}$ is proportional to $\sigma^4$.

It is interesting to note that in the region to the left from the
point $(kR_c)_{cr}$ we have observed that the length $L_n^{SB}$
for symmetric boundaries is twice as much as the length $L_n^{UB}$
for the uncorrelated boundaries. On the other hand, the latter
length is twice as much as the length $L_n^{AB}$ for the
antisymmetric boundaries, i.e., $L_n^{SB}\approx 2L_n^{UB}\approx
4L_n^{AB}$.

There is another border that is characterized by the transition to
the large-scale roughness with strong correlations, where a very
different behavior of the total inverse length $1/L_n$ was found
for the antisymmetric boundaries, in comparison with the
uncorrelated and symmetric ones. For the two last cases, the major
mechanism of the scattering is due to the amplitude terms in the
potential, and a similar behavior with respect to $kR_c$
($1/L_n\propto kR_c$) occurs for both of them. However, the
attenuation length for the uncorrelated boundaries turns out to be
twice as much as the length for the symmetric boundaries, i.e.,
$L_n^{UB}\approx 2 L_n^{SB}$ (note that this relation is inverted
in comparison with the region of small-scale roughness).

As for the waveguide with the antisymmetric boundaries for which
the amplitude terms are absent in the potential, the SGS-mechanism
prevails in this region, leading to the dependence $1/L_n\propto
(kR_c)^{-3}$. Note, that here there is the interplay between the
SGS-term and gradient one. The transition point was found to be
$kR_c\sim k\Lambda_n$, with the crossover specified by the
following interplays:
\begin{itemize}
\item
uncorrelated and symmetric boundaries: the amplitude and gradient
terms versus the amplitude terms,
\item
antisymmetric boundaries: the gradient and SGS terms versus
SGS-term.
\end{itemize}

Finally, there is an intermediate region of a large-scale
roughness and weak correlations where the first attenuation length
mainly contributes, $L_n^{(1)}\ll L_n^{(2)}$. Here one can observe
a strong similarity of the inverse attenuation length behavior for
all configurations considered. It should be noted that this region
is quite difficult for obtaining simple analytical estimates due
to a strong dependence of the number of terms in the corresponding
sums on the parameter $kR_c$. Numerical data manifest that the
attenuation length roughly remains the same inside this
intermediate region.

\begin{acknowledgments}
The authors acknowledge the support by the CONACYT (M\'exico) grant
No~43730.
\end{acknowledgments}



\end{document}